\documentclass[universe,review,accept,pdftex]{Definitions/mdpi} 

\usepackage{subfigure}
\usepackage{graphicx}
\usepackage{amssymb}
\usepackage{url}
\usepackage[italicdiff]{physics}
\usepackage{natbib}
\usepackage{xcolor}
\usepackage{siunitx}
\usepackage[nice]{nicefrac}
\setcitestyle{authoryear}

\firstpage{1} 
\pubvolume{xx}
\issuenum{7}
\articlenumber{7}
\pubyear{2021}
\copyrightyear{2021}
\history{Received: 31 March 2021; Accepted: 16 June 2021; Published: 24 June 2021}
\def\stacksymbols #1#2#3#4{\def\theguybelow{#2}
        \def\verticalposition{\lower#3pt}
        \def\spacingwithinsymbol{\baselineskip0pt\lineskip#4pt}
        \mathrel{\mathpalette\intermediary#1}}
\def\intermediary #1#2{\verticalposition\vbox{\spacingwithinsymbol
        \everycr={}\tabskip0pt
        \halign{$\mathsurround0pt#1\hfil##\hfil$\crcr#2\crcr
                \theguybelow\crcr}}}

\def\gta{\stacksymbols{>}{\sim}{2.5}{.2}}

\def\kms{{\rm km\,s^{-1}}}

\def\R500{$R_{500}$}
\def\r500{R_{500}}

\newcommand{\msun}{M_{\odot}}

\newcommand{\es}{{\rm erg\,s^{-1}}}

\newcommand{\Chandra}{\textit{Chandra}}
\newcommand{\Athena}{\textit{Athena}}

\newcommand{\XMM}{XMM-\textit{Newton}}
\newcommand{\ROSAT}{\textit{ROSAT}}
\newcommand{\ASCA}{\textit{ASCA}}
\newcommand{\Suzaku}{\textit{Suzaku}}
\newcommand{\XRISM}{\textit{XRISM}}
\newcommand{\Hitomi}{\textit{Hitomi}}
\DeclareMathSymbol{\la}{3}{AMSa}{46}

\Title{The metal content of the hot atmospheres of galaxy groups}

\Author{Fabio Gastaldello$^{1\,\orcidF{}}$, Aurora Simionescu$^{2,3,4,\orcidA{}}$, Fran\c{c}ois Mernier$^{5,2\,\orcidR{}}$, Veronica Biffi$^{6,7\,\orcidV{}}$, Massimo Gaspari$^{8,9\,\orcidM{}}$, Kosuke Sato$^{10\,\orcidS{}}$ and Kyoko Matsushita$^{11\,\orcidK{}}$}

\AuthorNames{Fabio Gastaldello, Aurora Simionescu, Francois Mernier, Veronica Biffi, Massimo Gaspari, Kosuke Sato, Kyoko Matsushita}

\address{%
$^{1}$ \quad IASF – Milano, INAF, Via A. Corti 12, I-20133 Milano, Italy \\
$^{2}$ \quad SRON Netherlands Institute for Space Research, Sorbonnelaan 2, 3584 CA Utrecht, The Netherlands \\ 
$^{3}$ \quad Leiden Observatory, Leiden University, PO Box 9513, 2300 RA Leiden, The Netherlands \\ 
$^4$ \quad Kavli Institute for the Physics and Mathematics of the Universe (WPI), The University of Tokyo, Kashiwa, Chiba 277-8583, Japan \\ 
$^{5}$ \quad European Space Agency (ESA), European Space Research and Technology Centre (ESTEC), Keplerlaan 1, 2201 AZ Noordwijk, The Netherlands\\ 
$^{6}$ \quad INAF - Osservatorio Astronomico di Trieste, via Tiepolo 11, I-34143 Trieste, Italy\\ 
$^{7}$ \quad IFPU - Institute for Fundamental Physics of the Universe, Via Beirut 2, I-34014 Trieste, Italy\\ 
$^{8}$ \quad INAF - Osservatorio di Astrofisica e Scienza dello Spazio di Bologna, via Piero Gobetti 93/3, I-40129 Bologna, Italia\\ 
$^{9}$ \quad Dept. of Astrophysical Sciences, Princeton University, 4 Ivy Lane, Princeton, NJ 08544, USA\\ 
$^{10}$ \quad Department of Physics, Saitama University, 255 Shimo-okubo, Sakura-ku, Saitama-shi, Saitama 338-8520, Japan\\
$^{11}$ \quad Department of Physics, Tokyo University of Science, 1-3 Kagurazaka, Shinjuku-ku, Tokyo 162-8601 Japan
}

\corres{Correspondence: fabio.gastaldello@inaf.it}

\abstract{
Galaxy groups host the majority of matter and more than half of all the galaxies in the Universe. Their hot ($10^7$ K), X-ray emitting intra-group medium (IGrM) reveals emission lines typical of many elements synthesized by stars and supernovae. Because their gravitational potentials are shallower than those of rich galaxy clusters, groups are ideal targets for studying, through X-ray observations, feedback effects, which leave important marks on their gas and metal contents. 
Here, we review the history and present status of the chemical abundances in the IGrM probed by X-ray spectroscopy. We discuss the limitations of our current knowledge, in particular due to uncertainties in the modeling of the Fe-L shell by plasma codes, and coverage of the volume beyond the central region. We further summarize the constraints on the abundance pattern at the group mass scale and the insight it provides to the history of chemical enrichment.
Parallel to the observational efforts, we review the progress made by both cosmological hydrodynamical simulations and controlled high-resolution 3D simulations to reproduce the radial distribution of metals in the IGrM, the dependence on system mass from group to cluster scales, and the role of AGN and SN feedback in producing the observed phenomenology. Finally, we highlight future prospects in this field, where progress will be driven both by a much richer sample of X-ray emitting groups identified with eROSITA, and by a revolution in the study of X-ray spectra expected from micro-calorimeters onboard XRISM and ATHENA.
}


\begin{document}


\section{Introduction}



Two major astrophysical discoveries have provided key answers to the fundamental question of the origin of the chemical elements in the past century: the discovery that stellar nucleosynthesis is responsible for the production of all the heavy elements from lithium to uranium \citep{Eddington:1920,Hoyle:1946,Burbidge:1957} and the detection of line emission due to highly ionized iron in the X-ray spectra of the intra-cluster medium (ICM) \citep{Mitchell:1976,Serlemitsos:1977}. The impact of these two discoveries was extraordinary. The first one demonstrated that all the elements (with the exception of hydrogen, helium and traces of lithium and berillium produced by the Big Bang nucleosynthesis) are forged in the cores of stars and in supernovae (SNe) and that when a star explodes as a supernova it enriches the surrounding interstellar medium with freshly created elements. The second one showed that galaxies lost part of their synthesized elements and that there has been a considerable exchange of chemical elements between stars, galaxies and the hot plasma surrounding them. 
This also means that the chemical elements trace the formation and evolution of structure which is shaped by the physical processes occurring on a very wide range of spatial scales, from the size of single supernova remnants to cosmological volumes.  

The improvements in the stellar and supernova nucleosynthesis theory and modelization \citep[e.g.][and references therein]{Nomoto:2013,Kobayashi:2020} established that the major astrophysical sources of the chemical elements are: i) Core-collapse supernovae (SNcc) and their massive progenitors ($\gtrsim 8-10 M_\odot$) synthesise most of the O, Ne, and Mg of the Universe and a considerable fraction of Si and S (collectively called $\alpha$ elements as they are the result of fusion process involving the capture of $\alpha$ particles); ii) Type Ia supernovae (SNIa), whose progenitors are generally believed to be exploding white dwarfs in binary systems, synthesise Ar, Fe and the other Fe-peak elements such as Cr and Ni, and the remaining fraction of Si and S; iii) Asymptotic giant branch (AGB) stars produce mainly C, N which are ejected through stellar winds.

In astrophysics the term "metals" refers to all the elements heavier than helium, in contrast to the terminology adopted in other scientific disciplines, in part because all these elements make up a small contribution in number and mass with respect to H and He. 
A considerable fraction of the metals (and most of the iron) do not reside in the galaxies of groups and clusters and they have been expelled in their surrounding X-ray emitting hot atmospheres (for the purpose of this review we will define the hot atmosphere of groups as intra-group medium, IGrM). Indeed the iron share, i.e. the ratio of iron in the hot atmosphere and the iron locked up in stars in the galaxies, ranges from 1 up to 10 \citep[see for recent measurements][]{Renzini:2014,Ghizzardi:2021}. Therefore the key question is to understand the main transport mechanisms responsible for that unbalance \citep[see for a more detailed discussion][]{Renzini:2008,Schindler:2008}. There are two broad categories of mechanisms: i) extraction by ram pressure stripping and galaxy-galaxy interactions; ii) ejection by galactic winds powered from inside the galaxies themselves either by SNe (stellar feedback) or by the supermassive black hole (SMBH) at their center (in the so-called active galactic nucleus, AGN, feedback). Other important processes redistributing the metals within the hot atmospheres are the central AGN uplift of metals \citep[see the reviews by][and the companion review by \citet{Eckert:2021}]{McNamara:2007,McNamara:2012,Fabian:2012,Gitti:2012} and sloshing, i.e. the offset of the bulk of central part of the hot atmosphere from its hydrostatic equilibrium in its gravitational potential and the subsequent oscillations that may broaden the original distribution at larger scales \citep[see the reviews by][]{Markevitch:2007,ZuHone:2016}. Another source of metals could be the diffuse stellar component not associated to any single galaxy but to the global halo of the cluster or group (also know as intra-cluster light, ICL) polluting the ICM and the IGrM in situ \citep{Sivanandam:2009}.

The purpose of this paper is to review the status of the metal abundance measurements in the IGrM and the progress made by simulations to reproduce and interpret those measurements. It is a companion of the other reviews in this series addressing the scaling relations of these systems \citep{Lovisari:2021}, the impact of AGN feedback \citep{Eckert:2021}, the overall insight provided by simulations \citep{Oppenheimer:2021} and the properties of the particular class of fossil groups \citep{Aguerri:2021}.

Groups of galaxies (which can be defined as objects with total masses $M_{500}$ in the range $10^{13}-10^{14} M_\odot$, though see \citet{Lovisari:2021} for the unavoidable ambiguity of the definition of a galaxy group) bridge the mass spectrum between L* galaxies and galaxy clusters. They are known to host a significant fraction of the number of galaxies in the Universe \citep[e.g.][]{Mulchaey:2000,Eke:2006}, they form in the filaments of the cosmic web and not only in the nodes \citep[e.g.][]{Tempel:2014}, and they are bright enough to be relatively easily observable in X-rays, while having low enough masses such that complex baryonic physics (e.g. cooling, galactic winds, AGN feedback) begins to dominate above gravity, making these objects more than simple scaled down versions of galaxy clusters \citep[e.g.][and the above mentioned reviews in this series]{Ponman:2003,McCarthy:2010}.
Groups of galaxies appear as critical systems to understand the process of structure formation, the dynamical assembly of baryons in the dark matter halos, and the complex physical processes affecting both the gas and the stellar components. For all the above reasons it is important to study metals in groups: if for example it is more controversial with respect to clusters that they can be considered as "closed box", they can provide key information on the processes resulting in the redistribution and loss of metals and it is clearly instructive to investigate the metal budget in different types of systems \citep[see the interesting discussion in the review by][]{Maiolino:2019}.

Many excellent reviews exist already focusing either
on the more general topic of X-ray spectroscopy, mainly of the ICM, or directly on metal abundances both observationally and theoretically \citep{Renzini:2004,Renzini:2008,Werner:2008,Schindler:2008,Borgani:2008,Bohringer:2010,Kim:2012,dePlaa:2013,Mernier:2018Review,Biffi:2018R} that ease our work which will then focus on the topics more directly related to the recent updates about metal abundances in galaxy groups. In this respect the only reviews of the topic are about the early history of the metal abundance measurements in the IGrM described in \citep{Mulchaey:2000} and a more recent update including the \Chandra\ and \XMM\ data can be found in \citep{Sun:2012}. A comparison of the metal budget in groups compared to the one in clusters and elliptical galaxies has been discussed in \citep{Mernier:2018Review}.

The review is organized as follows. In \S\ref{s:obs}, we review the observational measurements, from the early \ROSAT\ and \ASCA\ results to the more recent \Chandra, \XMM\ and \Suzaku\ CCD and high spectral resolution (with the RGS instrument on board \XMM) results.  In \S\ref{s:theory} we discuss the theoretical framework and the insight from numerical simulations. In \S\ref{s:tele}, we discuss the most relevant upcoming missions which will provide a key contribution to the field and in \S\ref{s:conclusions} we present our final remarks.




\section{X-ray observations} \label{s:obs} 


 \subsection{The observational signatures of metals in the IGrM}\label{s:Fe-L}

The X-ray spectra of the hot, diffuse gas that fills the dark matter halos of galaxies, galaxy groups, and galaxy clusters, is typically described as an optically thin plasma in collisional ionisation equilibrium (CIE), composed mainly of primordial hydrogen and helium gas but containing trace amounts of heavier elements from C up to Ni. These approximations usually provide a sufficient description of the bulk of the emission, although subtle effects due to various deviations from a simple thermal model can sometimes become relevant -- we refer the reader to \citet{Gu:2018} for a review.

There are two noteworthy differences between the X-ray emission from the ICM and IGrM. Firstly, in the hot ICM of clusters of galaxies, the free-free bremsstrahlung continuum is the dominant radiation process \citep[see e.g.][]{Mewe:1986}. Conversely, for plasma temperatures around and below 1~keV, which are typical of galaxy groups, increasing contributions to the continuum level come also from (i) recombination radiation, caused by the capture of an electron by an ion, leading to a spectral shape characterized by sharp ionization edges, and (ii) the slow transition from the 2s to the 1s state, which is forbidden by angular momentum conservation but can happen as a very slow two-photon process giving rise to continuum emission \citep[see e.g. Fig.6 of][]{Bohringer:2010}.
This makes it more difficult to determine the equivalent width of a given spectral line (EW\footnote{$EW=\int \nicefrac{(I_\nu-I_\nu^0)}{I_\nu^0} \:{\rm d}(h \nu )$, where $I_\nu^0$ is the continuum intensity without the line}), a quantity that serves as a main diagnostic of the abundance of the chemical element from which that line originates.

Secondly, at higher plasma temperatures ($\sim$4 keV and above), abundance measurements are typically driven by the signal obtained from the Fe XXV He-$\alpha$ line at 6.7~keV (rest frame). By comparison, below temperatures of around 2~keV, i.e. in the low-mass cluster and group regime, the most prominent diagnostic of the metallicity comes instead from the Fe-L line complex at 0.7--1.2 keV. This blend of emission lines originating from the L-shell transitions of Fe XVII -- Fe XXIV is completely unresolved at the spectral resolution of CCD cameras; the lines are so closely spaced together that, in many cases, they remain blended even for the \XMM\ RGS and the upcoming high-spectral resolution microcalorimeter onboard $XRISM$. Next to the dominant Fe-L lines, emission from Ne X and the L-shell of Ni at similar energies is also blended within the same spectral structure. 


In practice, elemental abundances are usually estimated by fitting the X-ray spectra with models of CIE emitting plasma, that account for the presence of various emission lines within the Fe-L blend (as well as the recombination and two-photon continua). The two plasma radiation codes most commonly used today in X-ray astronomy are AtomDB \citep{Smith:2001,Foster:2012} and SPEXACT \citep{Kaastra:1996,Kaastra:2020}. 
These models have evolved considerably over the last 4 decades, since the Fe-L emission was first discovered with the Solid-State Spectrometer onboard the $Einstein$ satellite \citep{Mushotzky:1981,Lea:1982}. 

 \begin{figure}[h]
 \centering
     \includegraphics[width=\textwidth]{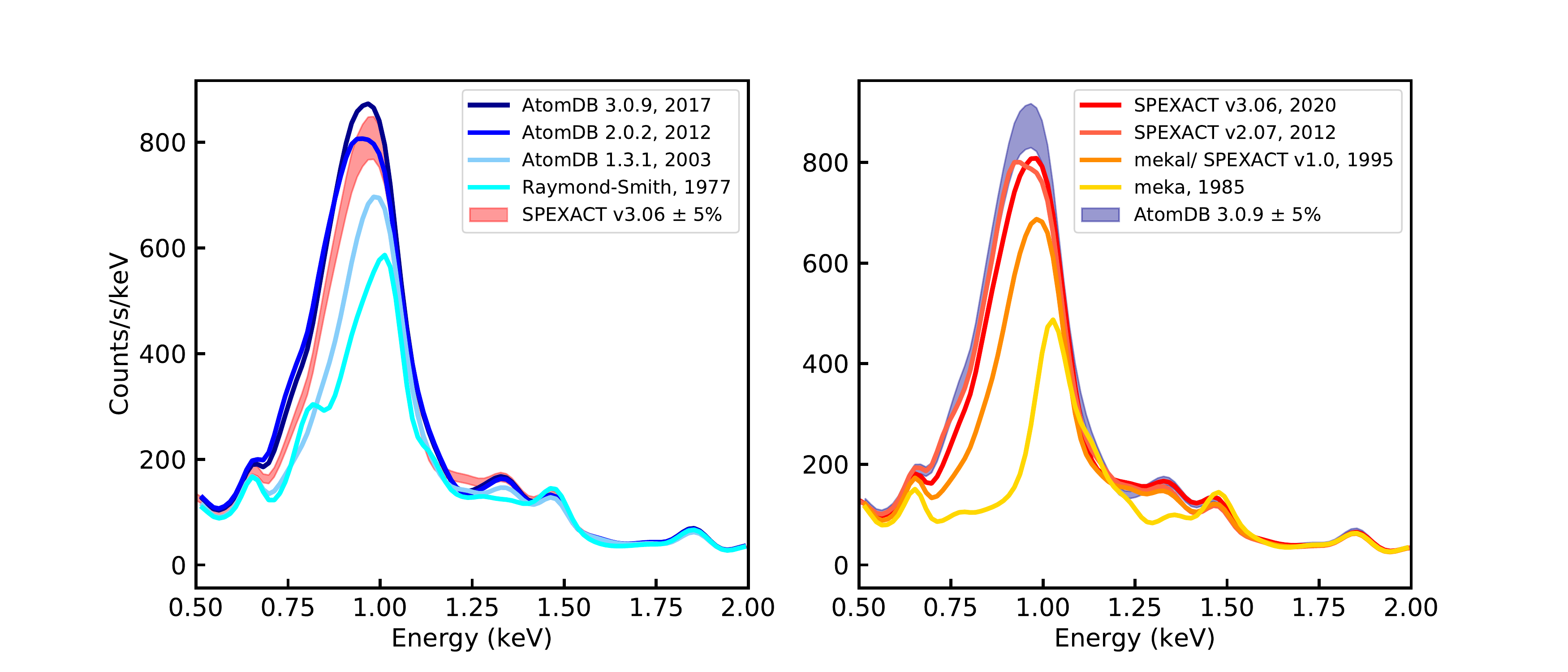}
     \caption{Historic evolution of the two most commonly used plasma emission models. For illustration, we focus on the specific case of a 1~keV plasma with Solar composition (following the Solar abundance units of \citealt{Asplund:2009}). All models have the same emission measure $\int n_e n_H dV$ and have been folded through the XMM-Newton MOS detector response.
     \textit{Left:} AtomDB \citep{Foster:2012} can be viewed as the replacement for / development of the Raymond-Smith code \citep{Raymond:1977}. \textit{Right:} SPEXACT originated from the `meka' model developed by Rolf \textbf{Me}we and Jelle \textbf{Ka}astra, which later became the `mekal' model; the addition of the final `l' comes from Duane \textbf{L}iedahl who calculated the atomic parameters for a large number of Fe L-shell ions \citep{Mewe:1985,Mewe:1986,Liedahl:1995}.}
     \label{fig_fel}
 \end{figure}
 
In Fig. \ref{fig_fel}, we illustrate the historic development of these two model `flavors' over time, using the specific example of a 1~keV plasma with Solar elemental composition, folded through the spectral response of a CCD camera. Simply and perhaps simplistically put, transitions belonging to lower ionisation states of Fe (whose emission lines lie at lower energies within the Fe-L `bump') require more complex computations and were therefore not fully accounted for in earlier models. This is why, both for AtomDB and SPEX, the shape of the lower energy wing of the Fe-L blend in particular appears to evolve significantly, and great care must be taken when discussing results obtained with older precursors of these plasma emission codes. 
It is encouraging that the models seem to be converging in recent years and, at least at CCD-level spectral resolution, the latest versions of AtomDB and SPEX only differ at about the $\sim5-10$\% level for a 1~keV plasma. Nevertheless, larger differences still remain for lower temperatures (up to 20\% at kT$<0.4$~keV when folded through a CCD response), and when viewing the models at higher spectral resolution, where the brightest lines in the Fe-L complex can be distinguished from the blend \citep[for a general and up-to-date discussion of the discrepancies of the two plasma codes see the discussion in][]{Mernier:2020a}. 

\begin{figure}[t]
 \centering
 \includegraphics[width=0.8\textwidth]{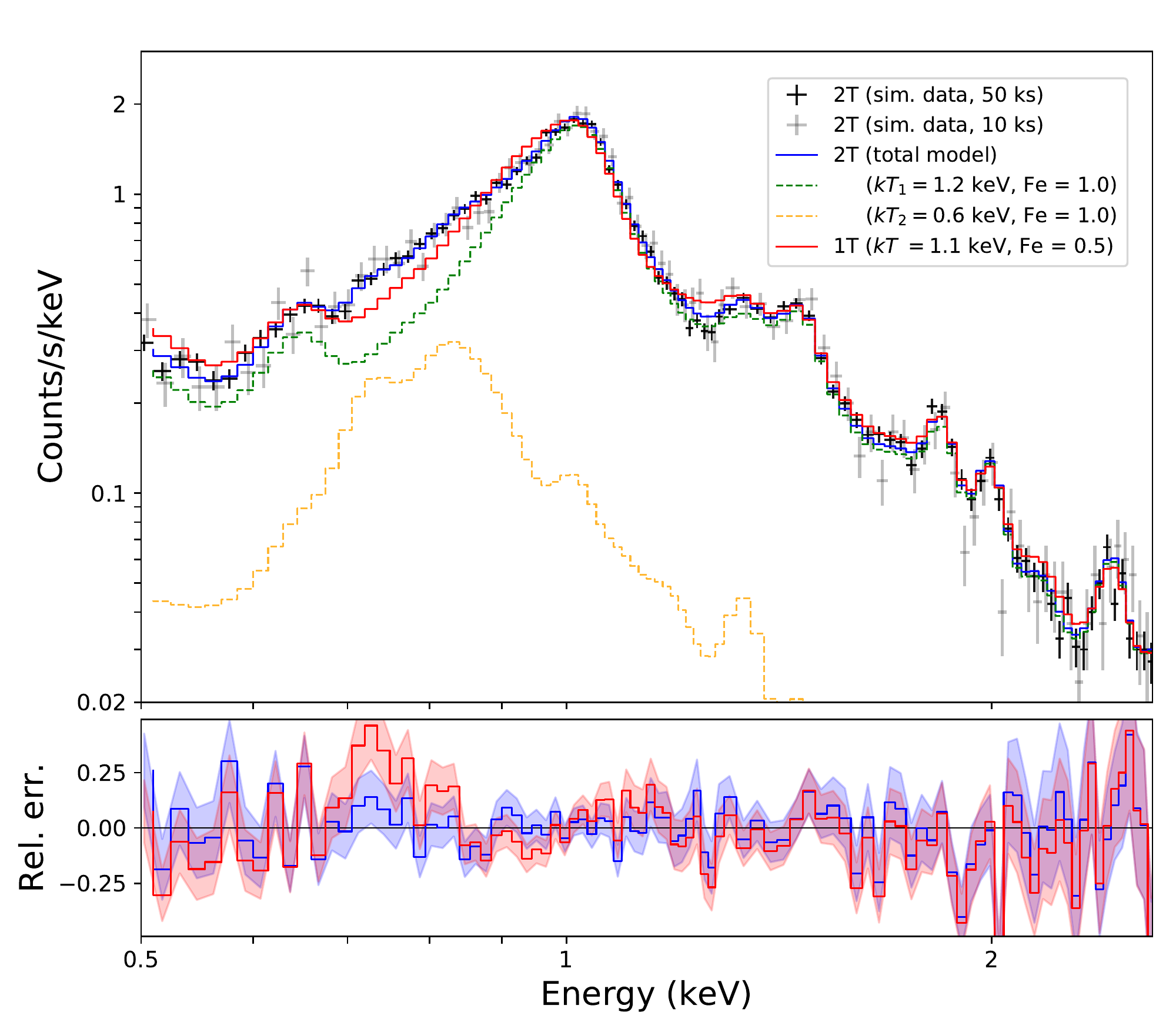}
      \caption{Simulated \XMM\ MOS observations of a multi-temperature plasma, illustrating the effect of the Fe bias. A mix of 0.6 and 1.2 keV plasmas with emission measures in a proportion of 1:10, Solar abundances in the units of \citet{Asplund:2009}, and a total flux similar to that of NGC5846 (integrated within 0.05r$_{500}$), was simulated using SPEXACT v3.0.6. Despite the fact that the low temperature component has only 1/10th of the emission measure of the hotter gas, a single temperature fit results in a best fit metallicity that is half of the value input to the simulation. The bottom panel shows the fit residuals for a 1T and 2T model for the 10~ks observation; while there is still a hint that the 1T model does not perfectly describe the data (given the positive residuals around 0.7 keV), this could easily be missed for fainter/more distant targets or when using smaller extraction regions for creating radial profiles or maps.}\label{fig_febias}
 \end{figure}

Observationally speaking, the Fe-L complex is both a blessing and a curse. On one hand, as mentioned above, it involves modelling the emission of Fe ions with many remaining electrons, which is significantly more difficult than H- or He-like transitions. On the other hand, because these lines are very bright, the total 0.5--2.0 keV flux of a kT=0.6--0.8 keV IGrM plasma with a Solar composition is \textit{more than three times brighter} than a 4~keV ICM plasma with the same emission measure (defined as $\int n_e n_H dV$). Quite literally, metals make the IGrM shine, and emission from Fe in particular is crucial for detecting what would otherwise be extremely faint and diffuse plasma in galaxy groups. However, this also means that, unlike the case of the ICM, a knowledge of the metallicity in the IGrM is indispensable for converting the observed X-ray flux into a particle number density. Furthermore, since the total flux and shape of the Fe-L bump are extremely sensitive to the plasma temperature (e.g. see the difference between the Fe-L model for a 0.6 vs. a 1.2 keV plasma in Fig. \ref{fig_febias}), significant biases arise when a spectrum containing multiple temperature components (for instance due to projection effects or radiative cooling in the core of a group) is approximated by a single-temperature model. This effect, dubbed `the Fe bias', is discussed in detail already in \cite{Buote:1998} and \cite{Buote:2000b} using ASCA data of elliptical galaxies and galaxy groups. 
In that work, and many subsequent references thereto, it is consistently shown that, if the abundance measurements are driven by the Fe-L signal, a two-temperature model can yield best-fit metallicities more than twice higher than a single temperature approximation. Although this conclusion was originally reached with old versions of the Fe-L plasma emission, it remains true today, as we illustrate in Fig. \ref{fig_febias}. 
It is also worth mentioning in passing that, at least at CCD spectral resolution, multi-temperature models can only be constrained if the abundances of the two components are coupled to each other, which need not be true in nature. In addition, beside the IGrM being intrinsically multi-phase, the unresolved emission from low-mass X-ray binaries (LMXB) may be an important spectral component, at least near the center of the brightest group galaxy (BGG) \citep[e.g.][]{Irwin:2003}; if unaccounted for, this may lead to biases in the measured abundances as well. 

As a last cautionary note in terms of interpreting various metal abundance measurements quoted in the literature, it is important to remember that these are customarily reported with respect to the Solar number ratio of that element compared to H; however, this reference point, too, has evolved over the past few decades. While the Solar photospheric units of \cite{angr} are still the default in the Xspec fitting package, and therefore widely used, this reference value for Fe/H is between 43 and 48\% higher than reported by the more recent work of \cite{Lodders:2009} and \cite{Asplund:2009}, respectively. Hence, the Fe abundances reported by various groups can be considerably different depending on the Solar units assumed, and care must be taken when comparing the results. Here we choose to normalize all quoted abundances to the units of \cite{Asplund:2009}. 

Moreover, the absolute values of e.g. O/H (often assumed to be Solar, with the implied variations/uncertainties just stated above) also affect the way that absorption edges from our own Milky Way influence the spectrum; since most of the emission from the IGrM is in the soft band, using the correct Galactic $n_{\rm H}$ \citep[see for example the discussion in][]{Lovisari:2019} as well as the correct chemical composition of the absorbing gas is important for obtaining a robust characterization of the spectral properties of the IGrM.

Armed with this overview of the spectral characteristics and potential pitfalls of modeling the IGrM, in the following sub-sections we discuss how our observational picture of metals in galaxy groups has evolved over the past several decades. 

\subsection{The early measurements of global metallicity}\label{s:obs_hist} 

\begin{figure}[h]
 \centering
     \includegraphics[width=0.8\textwidth]{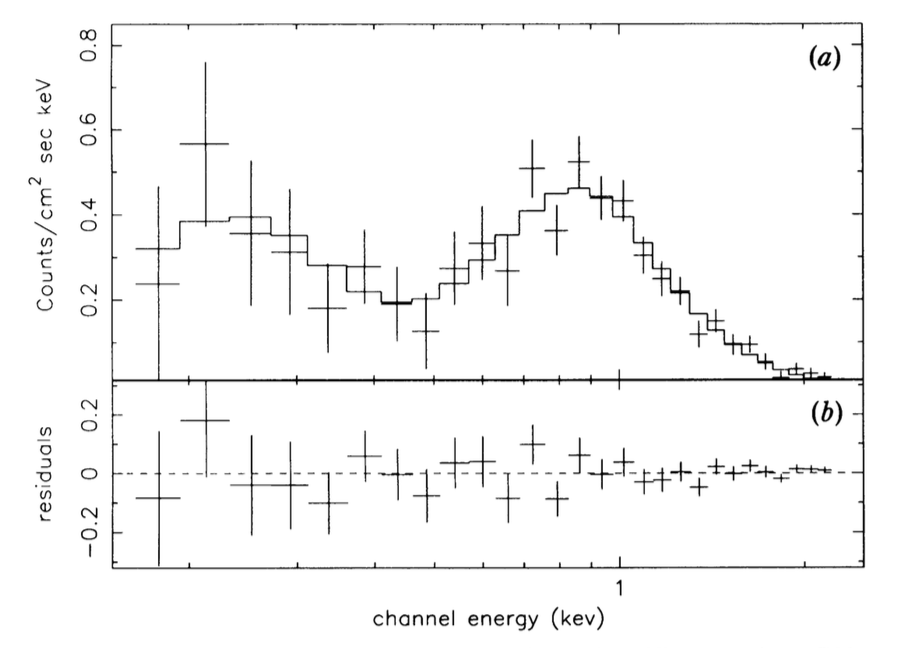}
     \caption{The \ROSAT\ X-ray spectrum of the group NGC 2300 plotted together with the beat-fit Raymond-Smith model (a) and residuals from the best fit model (b). Figure reproduced with permission from \citep{Mulchaey:1993}.}
     \label{fig_2300}
 \end{figure}

The first milestone discoveries of a true detection of the IGrM in the two galaxy groups NGC 2300 \citep{Mulchaey:1993} (see Fig.\ref{fig_2300}) and HCG 62 \citep{Ponman:1993} done with \ROSAT\ pointed to a surprisingly low abundance of the plasma of 0.09 Solar for NGC 2300 and 0.22 Solar for HCG 62, assuming the RS thermal plasma model \citep{Raymond:1977}. The entire detectable extent of the emission was fitted in a single aperture (25$^\prime$ for NGC 2300 and 18$^\prime$ for HCG 62, in the case of the analysis of NGC 2300 excluding the emission around the central galaxy itself).
 The \ROSAT\ observation of NGC 5044 \citep{David:1994} made it possible to measure spatially resolved temperatures and abundances, with super-Solar abundances in the inner 6$^{\prime}$ and beyond that radius consistent with a uniform distribution of 1.2 Solar. Very low abundance, $<0.12$, was reported in the NGC 4261 group from a fit with a RS model to a spectrum extracted from 40$^{\prime}$ \citep{Davis:1995}. In the first sample of HCGs \citep{Saracco:1995} a solid detection of extended emission in 3 of them (the already mentioned HCG 62, HCG 92 and HCG 97) and a possible detection in another 3, were all well fitted by a RS model with very low metal abundances. Additional \ROSAT\ observations allowed spatially resolved measurements for NGC 2300 confirming the low abundance, less than 0.16 Solar \citep{Davis:1996}. In a more complete survey of 85 HCGs with either deeper pointed (in 32 cases) or survey \ROSAT\ PSPC observations, extended emission from an IGrM was detected in 22 of them, including in the group emission also the emission spatially located on the dominant central elliptical \citep{Ponman:1996}. The metallicity derived for the 12 spectra of enough quality within an aperture of 200 kpc (in their cosmology) all pointed to a low abundance with a weighted mean of 0.27 Solar. Interestingly enough \citet{Ponman:1996} commented to treat these results with caution because the inferred low metal abundances rely heavily on the isothermal assumption: when temperature variations in the gas are taken into account, metallicities several times higher can be inferred. \cite{Mulchaey:1996} extended the search for \ROSAT\ observation of galaxy groups beyond HCGs with other optical catalogues in a sample of additional 14 groups finding emission from 4 of them and extending the census of the IGrM to 25 of the 48 groups analyzed at that time. Some conclusions were starting to be made, with the general lower abundance of the IGrM with respect to the ICM, despite the more equal share between gas and stellar mass, possibly suggesting that the IGrM may be largely primordial.

However concerns were starting to increase about the ability to model the dominant Fe-L emission in the IGrM by the available plasma codes. The first \ASCA\ CCD spectra (with the SIS instrument) of the cores of cool core clusters, Perseus, Abell 1795 and the Centaurus cluster exposed the limitations of both the RS and MEKA \citep{Mewe:1985,Mewe:1986} models \citep{Fabian:1994}, causing a major revision of the modeling of the Fe-L shell emission \citep{Liedahl:1995} which later was incorporated in the MEKAL code. These concerns
were reinforced by the discrepancy between the low metallicity found also in the inter-stellar medium of elliptical galaxies and the super-Solar abundances expected just by stellar mass loss \citep{Renzini:1993,Arimoto:1997}.

\ASCA\ measurements also reported a great scatter in the metallicity of the IGrM. The \ASCA\ study of NGC 5044 and HCG 51 reported metal abundances significantly higher than those of NGC 2300 and HCG 62 (also performed with \ASCA) and more similar to clusters \citep{Fukazawa:1996}.
\cite{Davis:1999} analyzed \ASCA\ data for 17 groups with single apertures ranging from 4$^{\prime}$ to 30$^{\prime}$ finding in general low abundances in the range 0.15-0.6 Solar.
The higher temperature and mass objects with \ASCA\ measurements reported by \citep{Hwang:1999}
showed an average abundance of 0.44 Solar consistent with that observed in rich clusters and therefore clearly highlighting the 1 keV regime as the one showing the spread to lower values of the abundance measurement.

The overall summary as done by \citet{Mulchaey:2000} is that of a surprising scatter in the measured metallicities in groups, from low (0.15-0.3 Solar) to higher than the values determined in clusters in those days (0.7-0.9) with \ROSAT\ and \ASCA.

A key insight was provided by a series of papers showing the biases introduced by fitting
with a single isothermal model complex spectra with multi-temperature components.
In the spectra extracted from large apertures in the bright cores of galaxy groups and ellipticals temperature and abundance gradients are present \citep{Buote:1998,Buote:1999,Buote:2000a,Buote:2000b}. The very sub-Solar abundances obtained from previous studies were an artefact of fitting isothermal models and two-temperature models provided better fits to the data and higher metallicities.
This is the Fe-bias also described in the previous section and demonstrated by means of simulations of \ASCA\ spectra \citep{Buote:2000b}. The discovery of the Fe bias highlighted the importance of the ability of performing spatially resolved spectroscopy and the difficulties in the modelization of the thermal and abundance structure in the cores of galaxy, groups and clusters, as it was shown at those times by the early results of M87 with \XMM\ \citep{Molendi:2001b}.

The last influential paper dealing with single measurements of metal abundances is  \citet{Baumgartner:2005}, presenting an analysis of the \ASCA\ spectra of 273 groups and clusters with the largest possible aperture collecting all the detectable flux and stacked in bins of temperature. That work found a constant Fe abundance value of 0.3 Solar for hot clusters and for groups with an increase up to a factor of 3 with respect to the average value in the range 2-4 keV.
This is a manifestation of the "inverse" Fe-bias \citep[][]{Rasia:2008,Simionescu:2009,Gastaldello:2010} which overestimates the abundances in multi-temperature plasma (ranging from about 1-2 keV to about 5 keV) resulting in a mean global temperature in the range 2-4 keV as found by \citet{Baumgartner:2005}. Although in practice when this occurs the spectra show the presence of both Fe-L and Fe-K lines, the inverse Fe-bias is essentially weighted by the higher statistics of the Fe-L complex. In this regime the fitting procedure increases the estimated Fe abundance to overcome the weaker Fe-lines expected in the single temperature plasma \citep[for more details see][]{Gastaldello:2010}.

\subsection{The spatial distribution of the metals in the IGrM}\label{s:spatial_distr}

Succeeding the \ROSAT\ and \ASCA\ eras, which allowed to set a first light on global metallicities of the IGrM, the advent of CCD instruments offered by the generation of early 2000's X-ray observatories (\Chandra, \XMM, \Suzaku) brought a significant progress. They allowed not only to reveal the spatial distribution of metals across galaxy groups, but also to focus on elements other than Fe -- hence exploring groups' chemical history with respect to its SNIa \textit{and} SNcc components. In the following subsections, we tackle these two aspects in more detail.

\subsubsection{Radial profiles of iron abundance} \label{s:radial} 

\begin{figure}[h]
 \centering
     \includegraphics[width=\textwidth]{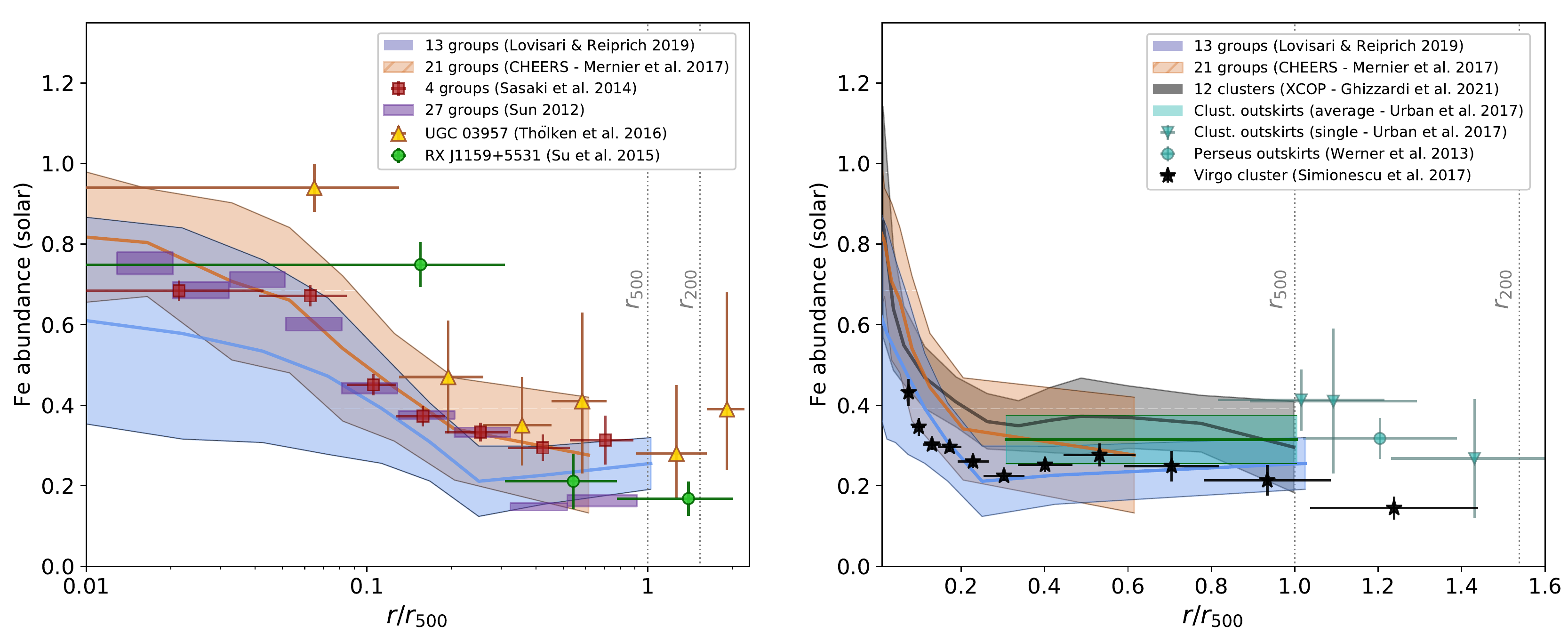}
     \caption{Fe abundance radial profiles in various galaxy groups (and clusters) from the literature. \textit{Left:} The recent average profiles of \citet[][the 21 CHEERS groups]{Mernier:2017} and \citet[][13 groups -- excluding systems overlapping with the CHEERS observations]{Lovisari:2019} are compared with those of \citet[][4 groups]{Sasaki:2014} and \citet[][27 groups -- below $kT_{500} = 1.9$~keV]{Sun:2012} as well as with independent measurements of UGC\,03957 \citep{Thoelken:2016} and RX\,J1159+5531 \citep[][azimuthally averaged]{Su:2015}. \textit{Right:} The same two average group profiles are compared with measurements of more massive systems -- i.e. the XCOP sample \citep[][12 clusters]{Ghizzardi:2021}, the Fe abundance in cluster outskirts (\citealt{Urban:2017}; with averaged and single measurements shown respectively below and beyond $r_{500}$ -- also including the outermost Perseus value from \citealt{Werner:2013b}), as well as the (azimuthally averaged) profiles of the Virgo cluster \citep{Simionescu:2017}. For consistency, the scatter envelope of the samples of \citet{Urban:2017} and \citet{Lovisari:2019} have been computed following that of the CHEERS sample \citet{Mernier:2017}. All measurements have been rescaled into radial units of $r_{500}$ \citep[following the values given in the corresponding papers and/or the conversion proposed by][]{Reiprich:2013} and into Solar units of \citet{Asplund:2009}.}
     \label{fig_radial}
 \end{figure}

The essence of this subsection is summarised in Fig.~\ref{fig_radial}, where we show a few recent radial metallicity (i.e. Fe) profiles of galaxy groups (from both individual and sample measurements), with comparison with typical cluster profiles. These profiles, among with a number of other ones reported in the literature, are further discussed below. 

Radial metallicity profiles of individual sources have been in fact investigated by many authors using either \XMM, \Chandra, or \Suzaku\ \citep[or even earlier with \ROSAT;][]{Buote:2000b}. This is the case for systems such as NGC\,5044 \citep{Buote:2003,Sasaki:2014}, RX\,J1159+5531 \citep{Humphrey:2012,Su:2015}, AMW\,4 \citep{O'Sullivan:2005a}, HCG\,62 \citep{Rafferty:2013,Sasaki:2014,Panagoulia:2015,Hu:2019}, MKW\,4 \citep{O'Sullivan:2003b,Sasaki:2014}, NGC\,1399 \citep{Buote:2002,O'Sullivan:2003a,Su:2017}, and UGC\,03957 \citep{Thoelken:2016}. 

The vast majority of these studies show a gradual metallicity increase towards the core of the systems, with the maximum value spanning from half a Solar to slightly super-Solar values. This picture is qualitatively in line with the centrally peaked Fe abundance profiles that are typically found in relaxed clusters \citep[e.g.][]{DeGrandi:2001,Leccardi:2008b,Mernier:2017,Lovisari:2019,Ghizzardi:2021}. Quantitatively, samples including groups \textit{and} clusters are valuable to provide comprehensive comparisons. For instance, \citet{Johnson:2009,Johnson:2011} studied 28 galaxy groups and concluded that (i) systems with lower level of feedback impact are on average more metal rich within $\sim$0.03$R_{500}$, and (ii) systems classified as "cool-cores"\footnote{In \citet{Johnson:2009}, a "cool-core" group is defined as such when its temperature profile shows a clear central decrease out to at least $\sim$0.1$R_{500}$. This definition is of course arbitrary and may differ from other proposed ones.} are, on average, more enriched in their cores ($\sim$0.1$R_{500}$) than clusters. Similar conclusions were reached for 43 \Chandra\ groups (re-) analysed by \citet{Sun:2012}, although a significant increase of average metallicity with group temperature (hence, mass) was also reported. More recently, results from the CHEERS sample -- consisting of 21 "groups/ellipticals" and 23 "clusters"\footnote{In \citet{Mernier:2017}, a "group/elliptical" is defined as such when the mean temperature of the system within 0.05$R_{500}$ does not exceed 1.7 keV. Here again this definition is arbitrary and a given rich group could be defined as a poor cluster in other studies.} -- suggest for instance a similar decreasing profile for both types, with the former being on average slightly less enriched than the latter \citep{Mernier:2017}. On the other hand, \citet{Lovisari:2019} analysed a sample of 207 systems and concluded that, despite their scatter, on average "groups/ellipticals" (defined in the same way) have a slightly higher metallicity than clusters within 0.1$R_{r500}$. A recent re-analysis of the CHEERS sample within $0.1R_{500}$ using an updated SPEX version (v3.0.4, also more consistent with the apec v3.0.9 version used in \citet{Lovisari:2019}) find more consistent results, with groups being at least as enriched as clusters within that limit \citep{Mernier:2018a}. More detailed discussions and interpretations on the absolute metallicities in groups. vs. more massive systems is addressed in Sect. \ref{s:mlr} (observations) and Sect. \ref{s:cosmo-res} (cosmological simulations). 

Also quite remarkably, Fig. \ref{fig_radial} suggests that the sample-averaged metallicity gradient measured from these different authors all have a similar slope. One notable exception (not shown here) is perhaps the sample of 15 nearby groups observed with \Chandra\ by \citet{Rasmussen:2007,Rasmussen:2009}, whose average profile exhibits a significantly sharper central peak \citep[as already pointed out by][]{Sun:2012}. This difference might originate from spectral modelling (including outdated atomic data and/or multi-temperature biases), instrumental calibration, or subtle background effects, which were all less understood at that time.
 
It is worth noting that the metallicity does not always increase with decreasing radius. In fact, for a number of systems, the Fe abundance was found to peak a few kpc outside of the core while \textit{decreasing} towards its very centre. Although historically discovered and investigated in the Centaurus cluster \citep{Sanders:2002}, these drops are more commonly found in lower-mass systems \citep[e.g.][]{Rasmussen:2007,Rafferty:2013,Panagoulia:2015,Mernier:2017,Gendron-Marsolais:2017}. Whether the presence of these drops is truly related to mass of the system (and/or the strength of their cool core) is not clear yet. Indeed, abundance drop detections might be affected by selection biases -- originating from either the usually larger distance of clusters (resulting in a poorer spatial resolution, hence no detected drop), or the selection itself of the currently most studied systems.

Such low abundances are in fact surprising and intriguing, as they cannot be easily explained by classical models of IGrM formation and enrichment. Although in some cases drops were found to be the result of spectroscopic biases \citep[e.g. multi-temperature bias;][]{Werner:2006a}, no evidence points toward the latter as being the sole explanation. Similarly, resonant scattering seems to be excluded from the culprit list in at least a few specific cases \citep{Sanders:2006}, and possible helium sedimentation -- leading to an incorrect estimate of the continuum -- should provide limited effects only, if not largely inhibited by thermal diffusion \citep[][and references therein]{Ettori:2006,Medvedev:2014}. Interestingly, however, probing the \textit{chemical composition} in the central low-Fe regions of these systems may provide an interesting hint toward the physical nature of these drops. This is further discussed in Sect. \ref{s:composition}.

Another important open debate concerns the comparison of clusters' and groups' metallicities in their outskirts ($\sim R_{500}$ and beyond). Whereas there is now striking evidence for clusters having their metallicity flattening with radius and converging toward an universal value of $\sim$0.3 Solar \citep{Werner:2013b,Urban:2017,Ghizzardi:2021}, groups and elliptical galaxies have sometimes been measured with an uninterrupted decrease of metallicity down to at most 0.1--0.2 Solar \citep[e.g.][]{O'Sullivan:2003a,Buote:2004,O'Sullivan:2007,Su:2015}. The trend seems to be followed by the sample results of \citet{Rasmussen:2007,Rasmussen:2009} and of \citet{Sun:2012}. These results, however, should be interpreted with caution, given how recent atomic codes improvements changed our view on the Fe-L complex, its modelling at moderate resolution, and its associated abundance (Sect. \ref{s:Fe-L}). Moreover, some past measurements may have been affected by the Fe-bias discussed above, as moderate exposures available per source did not necessarily allow to model outer regions with more than one temperature. We note, for instance, that the more recent sample measurements of \citet{Mernier:2017} and \citet{Lovisari:2019} show hints of a flattening beyond $\sim 0.3 R_{500}$ that remains formally consistent with the 0.3 Solar value reported in clusters \citep[][see Fig. \ref{fig_radial}]{Urban:2017,Ghizzardi:2021}. These results are in agreement with \citet{Thoelken:2016}, who reported that the metallicity profile of the group UGC\,03957 does not decrease further below 0.3 Solar, even at distances \textit{beyond} $R_{200}$. Other measurements, on the other hand, show in-between results, with evidence of a flattening though around 0.2 Solar, i.e. \textit{below} the universal value. This is the case for the galaxy group RX\,J1159+5531 \citep{Su:2015}, as well as (perhaps even more intriguingly) for the Virgo cluster \citep{Simionescu:2017}.


The question of whether groups and clusters have their outskirts enriched at similar levels is of crucial importance. Besides the fact that outskirts represent by far the largest volume of these systems (hence the bulk of their metal masses), they are direct witnesses of freshly accreted gas through the gravitational potential of these systems and thus constitute a fossil record of the enrichment of these systems at their formation epoch. In fact, the (radially \textit{and} azimuthally) uniform metallicity distribution measured in clusters outskirts constitutes by far our best evidence in favor of an "early-enrichment" scenario, in which supermassive black hole feedback played a fundamental role in ejecting and mixing freshly produced metals out of their galaxy hosts during or before their assembly into larger scale structures and the formation of their hot ICM, i.e. at $z \gtrsim 2$--3 \citep[for recent reviews, see e.g.][]{Biffi:2018,Mernier:2018Review}. Quite remarkably, this redshift range also corresponds to the peak of star formation activity \citep[for a review, see e.g.][]{Madau:2014}, as well as to an epoch of enhanced AGN accretion and activity -- not only at cosmic scale \citep[for a review, see e.g.][]{Hickox:2018} but also (and especially) in clusters and groups \citep[e.g.][]{Martini:2013,Bufanda:2017,Krishnan:2017}, naturally leading to the picture of their higher feedback to efficiently stir the freshly produced metals. Robust measurements revealing a uniform metal distribution in the IGrM as in the ICM will constitute decisive evidence towards this scenario and its "universal" 0.3 Solar value. On the contrary, significantly lower abundances measured in the IGrM outside $\sim R_{500}$ would challenge this scenario and would require to rethink our global picture of chemical enrichment at galactic scales and beyond. High resolution spectroscopy coupled to high throughput will be essential to bring our current measurements up to the required accuracy (Sect. \ref{s:tele}).

\subsubsection{Chemical composition, and its radial dependence} \label{s:composition}

Since Fe has the strongest emission lines in the IGrM, it typically dominates the abundance measurements reported in the literature. For low-statistics spectra, it is common to assume that the abundances of other elements with respect to Fe follow the Solar ratio. However, important information about the metal enrichment history of the IGrM is encoded in its chemical composition, in particular since the O/Fe, Mg/Fe, and/or Si/Fe ratios are good tracers of the relative contribution of SNcc and SNIa. This relative contribution is expressed in various ways throughout the relevant literature, for example as the ratio between the numbers of different supernova explosions (either $N_{cc}/N_{Ia}$ or $f_{Ia}=\nicefrac{N_{Ia}}{N_{Ia}+N_{cc}}$); or as the fraction of Fe supplied by SNIa, $f_{\rm Fe,Ia}=N_{Ia}*y_{Ia,Fe}/(N_{cc}*y_{cc,Fe}+N_{Ia}*y_{Ia,Fe})$, where $y_{SN,i}$ represents the mass of element $i$ produced by a supernova of type $SN$.
The details of this decomposition depend on the exact model yields $y_{SN,i}$, which are subject to remaining uncertainties in stellar astrophysics, and furthermore rely on various assumptions about the initial metallicity and mass function of the supernova progenitors. Nevertheless, the general trend wherein light-$\alpha$ elements are almost exclusively produced by SNcc while Fe-group elements are mainly supplied by SNIa is robust among the current chemical evolution models, lending credibility to this type of analysis.


Back to the \ROSAT\ and \ASCA\ era, \citet{Finoguenov:1999} reported from a sample of four galaxy groups that SNcc products (i.e. Si and Mg) were found to be more uniformly spread, while SNIa products (namely, Fe) showed a more peaked distribution. These results were naturally interpreted as the bulk of SNcc having exploded, gotten mixed with, and enriched their surroundings earlier than the bulk of SNIa (the latter being more likely to originate from long-lived low-mass star populations in the red-and-dead central dominant galaxy). That interpretation was later supported by \citet{Rasmussen:2007,Rasmussen:2009} who measured a radial increase of the Si/Fe ratio with \Chandra\ observations of 15 groups. 

However, these initial conclusions do not appear to have stood the test of time. Some early \XMM\ data already provided results that conflicted with the initial paradigm of a relatively uniform $\alpha$-element and peaked Fe distribution in the IGrM: no gradient in Si/Fe was seen in NGC5044, comparing the regions within and beyond 48~kpc from the BGG \citep{Buote:2003}; a constant and close to Solar $\alpha$/Fe out to at least 100 kpc was reported in NGC507 \citep{Kim:2004}; and \citet{Xue:2004} found that \textit{all} measured abundances (O, Mg, Si, S, and Fe) in the group RGH80 showed a monotonic decrease with radius. In all these three cases, $f_{\rm Fe,Ia}$ was inferred to be in the range of 70--85\%, assuming a model consisting of simple linear combinations of SNIa and SNcc. Therefore, although most Fe is being supplied indeed by SNIa, there did not appear to be a significant change in $f_{\rm Fe,Ia}$ with radius, or from system to system. 
Similar conclusions were starting to be reached in galaxy clusters as well \citep[e.g. ][]{Gastaldello:2002,dePlaa:2006,Werner:2006a,Simionescu:2009}. 

The low instrumental background of \Suzaku, and the superior low-energy response of the XIS CCDs particularly in the first few years after launch, shed additional light on this topic: the radial profiles of Mg/Fe, Si/Fe, and S/Fe were consistently shown to remain uniform (i.e. all four elements showed a radially decreasing profile) in HCG 62 \citep{Tokoi:2008}, NGC 5044 \citep{Komiyama:2009}, NGC 507 \citep{Sato:2009}, and NGC 1550 \citep{Sato:2010} over the entire area probed by the \Suzaku\ observations. A sample of 4 groups consisting of MKW4, HCG62, NGC1550, and NGC5044, was covered by \Suzaku\ out to as far as 0.5 $r_{180}$, confirming that the Mg/Fe and Si/Fe ratios remain nearly constant and close to the Solar ratio \citep[assuming the units of][]{Lodders:2003} out to a significant fraction of the virial radius \citep{Sasaki:2014}. All these measurements are consistent with an $N_{cc}/N_{Ia}$ of 3--4 \citep{Sato:2007,Komiyama:2009,Sato:2010}. For the supernova yields assumed in these works, $N_{cc}/N_{Ia}=3$ corresponds to a $f_{\rm Fe,Ia}$ of 80\%, in line with the \XMM\ results discussed in the previous paragraph. 
A point of contention in the \Suzaku\ results remained the O abundance, that seemed to have much shallower radial gradients than all other $\alpha$-elements (a conclusion shared by all references mentioned earlier in this paragraph). This would imply increasing O/Mg and O/Si ratios as a function of radius; since all these three elements are predominantly produced by SNcc, it is impossible to reconcile these measurements with a simple model where the relative contribution from SNIa and SNcc varies with distance from the BGG. It is likely that residuals in modeling the Galactic absorption and/or OVIII foreground emission, or issues related to Solar Wind Charge Exchange (whose strongest emission also comes from O), may have affected the measurements.

More recently, results from the \XMM\ CHEERS sample reported similar radial distributions of the O, Mg, Si, S, Ar, Ca, and Fe abundances in clusters \textit{and} groups separately. While the covered radial range does not extend as far as that from \Suzaku\ studies, the agreement between the radial trends of all measured elemental abundances, together with the larger sample size, provides solid and cohesive evidence for a lack of significant spatial variation of the SNcc vs. SNIa contributions to the enrichment across the mass scale \citep{Mernier:2017}. 

 \begin{figure}[h]
 \centering
     \includegraphics[width=0.85\textwidth]{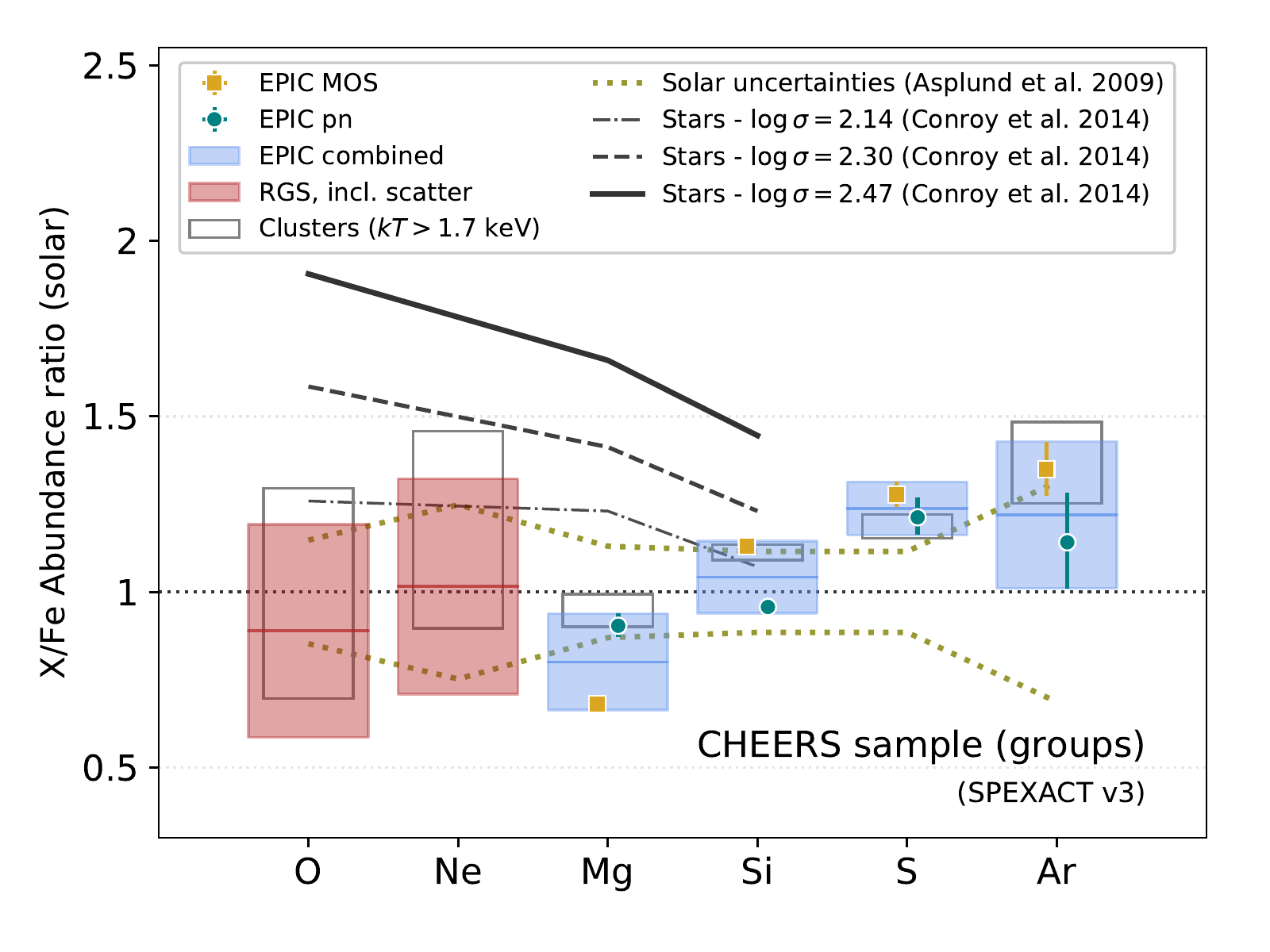}
     \caption{Average chemical composition \citep[expressed as X/Fe abundance ratios in units of][]{Asplund:2009} within the central 0.05$r_{500}$ of the 21 galaxy groups and massive ellipticals from the CHEERS sample (defined as $kT_\mathrm{mean} < 1.7$~keV). The O/Fe and Ne/Fe ratios (including their intrinsic scatter) were measured using the \XMM\ RGS instruments, while the other ratios are measured using the \XMM\ EPIC MOS and pn instruments. To be conservative, the EPIC "combined" measurements cover the entire MOS-pn discrepancies, which fully accounts for the intrinsic scatter as well. Comparison with the average chemical composition of clusters shows values that are similar and near-Solar in both regimes. For comparison, Solar uncertainties are also shown, as well as stellar abundances in ETGs measured for different bins of stellar velocity dispersion $\sigma$ \citep{Conroy:2014}. Adapted from \citet{Mernier:2018b}}.
     \label{fig_CHEERS}
 \end{figure}

With the latest advancements in our knowledge of spectral modelling, multi-temperature biases, and/or instrumental calibration, current measurements thus favour a uniform chemical composition over the entire volume of clusters and groups, as early suggested for the cores of both systems (and ellipticals) by \citet{DeGrandi:2009}. Interestingly, in both regimes the chemical composition is also remarkably close to that of our own Solar System. Indeed, \Hitomi\ confirmed that all the investigated X/Fe ratios of the Perseus Cluster are consistent with Solar at very high precision  \citep{Hitomi:2017,Simionescu:2019}, and detailed investigations of the CHEERS sample found the same trend for groups and ellipticals as well \citep{Mernier:2016a,Mernier:2018b}. This is further illustrated in Fig.~\ref{fig_CHEERS}, where we compiled the average chemical composition of the 21 CHEERS low-mass systems.

It is worth noting that the chemical composition of the ICM/IGrM must therefore be markedly different than that of the stars in the BCG/BGG: as shown by \cite{Johansson:2012,Conroy:2014}, massive early-type galaxies (ETGs) with a velocity dispersion above 200~km/s typically have high $\alpha$/Fe ratios up to twice the Solar value, which is inconsistent with the abundance pattern of the hot diffuse gas in their immediate vicinity (see Fig. \ref{fig_CHEERS}). High values of $\alpha$/Fe are usually associated with a very short starburst: BCG/BGGs may have made most of their stars before SNIa had time to explode, so that very few Fe-group elements are incorporated into the stars themselves. Nearly all SNIa later polluted the central ICM/IGrM instead, gradually lowering its $\alpha$/Fe ratios.

But although the SNIa contribution cannot have come too quickly (else the stars in the central galaxy would not have such a high $\alpha$/Fe), it also cannot have happened too slowly, or else the observed radial distribution of Fe should follow the present-day stellar light, which is not observed (Section \ref{s:mlr}). A significant late-time input of SNIa products would also modify the radial trends of $\alpha$/Fe in the ICM/IGrM, contradicting the constant near-Solar $\alpha$/Fe ratios measured throughout the volumes of clusters and groups.


This suggests that most (if not all) SNIa contributing to the enrichment exploded not much later than the peak of cosmic star formation \citep[$z \simeq 2$--3;][]{Madau:2014}. Several studies of the SNIa delay-time distribution in fact support this picture, finding that a significant number of such explosions occur as early as 100~Myr after a star formation event (\citealt{Totani:2008,Maoz:2012}; for a review on SNIa delay-time distribution and its interpretations, see \citealt{Maoz:2014}).

Nevertheless, the confirmation (or rejection) of this new paradigm will be crucial to achieve with future missions. Metal abundances determined with CCD spectrometers in clusters of galaxies are still subject to systematic uncertainties in the range of $\sim20$\% \citep{DeGrandi:2009,Simionescu:2019}; given the still ongoing challenge to derive accurate abundances from unresolved line complexes, these uncertainties may be even more important for the IGrM. Ultimately, the stellar population histories of typical central dominant galaxies are fundamentally different from that of the Milky Way. Future measurements using high-resolution spectroscopy will reach percent-level accuracy in determining the abundance of numerous chemical elements in gaseous halos of varying mass; it would be nothing short of a stunning cosmic conspiracy if, as smaller and smaller spatial scales start to be probed at such a level of precision, the central abundances in groups and clusters remain in agreement with the Solar composition.

Besides quantifying the relative contribution of SNIa and SNcc to the enrichment of the IGrM, elemental abundance ratios may also reveal the nature of the so far unexplained abundance drops that are sometimes observed (see \S \ref{s:radial}) in the very inner centers of groups and clusters. 
Under the assumption that these abundance drops have an astrophysical origin, an interesting scenario proposed by \citet{Panagoulia:2013,Panagoulia:2015} considers that IGrM-phase metals may deplete into dust and then become invisible to the X-ray window. As a second step, AGN jets and buoyantly rising bubbles may contribute to move this dust mass away, before eventually re-heating it to the X-ray phase outside of the very core. If true, an interesting corollary of this scenario concerns the Ne and Ar abundance. As these two elements are noble gases, they can not be incorporated into dust, hence they should \textit{not} exhibit any central decrease. Although a few authors have investigated this issue \citep{Mernier:2017,Lakhchaura:2019,Liu:2019Ar}, no real consensus is established yet given the sensitivity of the measurements to systematic effects. Ar and Ne lines should be easily measurable with future micro-calorimeters (Sect. \ref{s:tele}). Provided that atomic codes continue to converge (Sect. \ref{s:Fe-L}) in the years to come, \Athena\ (and possibly \XRISM\ for very nearby systems) will provide a definitive answer to this question.

Clarifying these outstanding issues will allow us to identify which combinations of theoretical supernova yield models, $y_{SN,i}$, provide the best fit to the observations of the ICM and IGrM (avoiding regions affected by dust depletion if applicable), offering important clues about open aspects of stellar astrophysics. For instance, it was realized very early on that abundance ratios measured from X-ray spectra of the ICM could be used to distinguish between various SNIa explosion mechanisms, preferring a deflagration over a delayed detonation model \citep{Dupke:2000}. Similar conclusions were tentatively reached for the IGrM \citep{Buote:2003,Sato:2007}. It is becoming clear, however, that both an improvement in the data quality and increased accuracy in the yield models are necessary before robust conclusions can be drawn (\citealt{DeGrandi:2009,Mernier:2016b,Hitomi:2017,Simionescu:2019}; for a review, see \citealt{Mernier:2018Review}). Significant progress is expected to be driven in this sense by upcoming high-resolution X-ray spectroscopy studies.


\subsubsection{2-Dimensional metallicity maps } 
\label{s:maps}

In addition to the radial dependence of the metal abundances, important information can be inferred also from the azimuthal substructure revealed by 2D maps; typically these are of course only available for very deep observations of the brightest systems. Early work by \citet{Finoguenov:2006} presented a systematic analysis of the metallicity distribution in  NGC5846, NGC4636, and NGC5044 using \XMM. It was shown that, while the profiles are consistent with a linear decrease with radius, the scatter of the data points was as high as 30-50\%. This pointed towards a patchiness of the 2D metal abundance using typical spatial resolution elements of 2 -- 10 kpc, which cannot be explained solely by the satellite subhaloes. Later studies revealed that, very generally speaking, the main physical mechanisms responsible for such a 2D metallicity substructure are related either to AGN feedback or to ongoing mergers. 

In terms of AGN feedback, in the case of clusters of galaxies, it is now well established that the buoyantly rising bubbles produced by the activity of the supermassive black hole in the BCG are able to uplift metals in their wake, leading to an abundance enhancement along the axis corresponding to the radio jets compared to the perpendicular direction \citep{Simionescu:2008,Simionescu:2009,Kirkpatrick:2011,Kirkpatrick:2015}. Given the shallower gravitational potential wells of galaxy groups, one might expect this effect to be even more pronounced, and even more important for the physical evolution of the IGrM, as metals produced in the BGG may even escape the group halo through the action of the AGN. However, to our knowledge, a systematic study of the metal asymmetry in groups \citep[i.e. an equivalent to the sample study of][ which focused primarily on galaxy clusters]{Kirkpatrick:2011,Kirkpatrick:2015} is still lacking. The main impediment is likely related to the fact that the region of uplift is also generally expected to be multi-phase \citep[see especially][]{Simionescu:2008} which, as discussed in \S \ref{s:Fe-L}, significantly complicates the determination of an exact Fe budget in a given spatial region. 

Nonetheless, hints that the relativistic radio lobes of the central AGN do have an impact on the metallicity distribution in groups have been obtained in a few objects. Perhaps the clearest example so far is that of AWM\,4, where \citet{O'Sullivan:2011} found a metal enhancement along the inner jet of the central dominant galaxy, NGC6051, corresponding to an excess mass of iron in the entrained gas of $\sim1.4\times10^6\:\rm{M}_\odot$. Another case is that of NGC4636, where \citet{O'Sullivan:2005b} report a plume of cool, metal-rich gas extending beyond a known AGN lobe to the southwest of the galaxy center, and interpret this to be the product of metal entrainment by past AGN activity. Less clear is the scenario in NGC4325; here \citet{Lagana:2015} reported an elongated Fe-rich filament to the south/southeast of the central galaxy which could be due to metal entrainment by the AGN; however, no X-ray cavity is found in this system that would confirm this interpretation. 
Finally, there are the interesting examples of rather clearly detected \textit{anti}correlations, i.e. a low metallicity corresponding to the radio lobes in NGC5813 \citep{Randall:2015} and inner radio lobes of M49 \citep{Gendron-Marsolais:2017} -- although, in both cases, only a single-temperature fit was used to create the 2D maps, hence the Fe bias may be the reason for these results. 

Mergers on the other hand typically result in tails and arcs of enhanced metallicity in the IGrM, depending on the merger stage and geometry. In a simple case where a subgroup is falling towards a larger cluster of galaxies, a ram-pressure stripped tail exhibiting an orderly head-tail morphology is often seen; as metals are stripped from the central group galaxy, the elemental abundances in this tail are expected to be higher than those of the surrounding diffuse medium. This has been confirmed by X-ray spectral mapping of the metallicity in a handful of cases. One of the clearest examples is that of the M86 group falling into the Virgo Cluster; this is a very rare case where a metal abundance map from a \textit{two-temperature} model is available for the IGrM \citep{Ehlert:2013}, showing a long, 100-150~kpc tail of near-Solar abundance (in units of \citealt{Grevesse:1998}) trailing M86. The abundance in the ram-pressure stripped tail is about twice higher than the off-tail regions, demonstrating how infalling groups contribute to the metal budgets of the ICM.  Another remarkable system is the northeastern group falling into Abell~2142, which exhibits a long, straight, narrow tail that flares out after about 300~kpc from the BGG. The metallicity map published in \citet{Eckert:2017_ramP} shows a significant enrichment along most of the narrow tail, with the transition between the straight tail and the irregular diffuse tail corresponding to a marked abundance drop. Recent spectral maps by \citet{O'Sullivan:2019} also show tails of cooler, lower entropy, metal-enriched gas behind both cores in a group-group (as opposed to group-cluster) merger in NGC 6338.

In later merger stages, after the first pericenter passage of the sub- and main halo, internal gas sloshing or tidal (also known as `slingshot') tails \citep{Hallman:2004,Sheardown:2019} can instead be recognized as arc-shaped high metallicity `fronts'. Internal gas sloshing is likely responsible for the high abundance arc in HCG~62 \citep{Gu:2007,Gitti:2010,Rafferty:2013,Hu:2019} and for the abundance map asymmetry in NGC5044 \citep{O'Sullivan:2014}; although no 2D metal abundance map is available, radial profiles of an azimuthally resolved wedge in the NGC 7618 -- UGC 12491 pair \citep{Kraft:2006} suggest a metal enhancement that was originally attributed to ram-pressure stripping but later recognized as rather due to a slingshot tail \citep{Sheardown:2019}.

 \begin{figure}[t]
 \centering
     \includegraphics[width=0.85\textwidth]{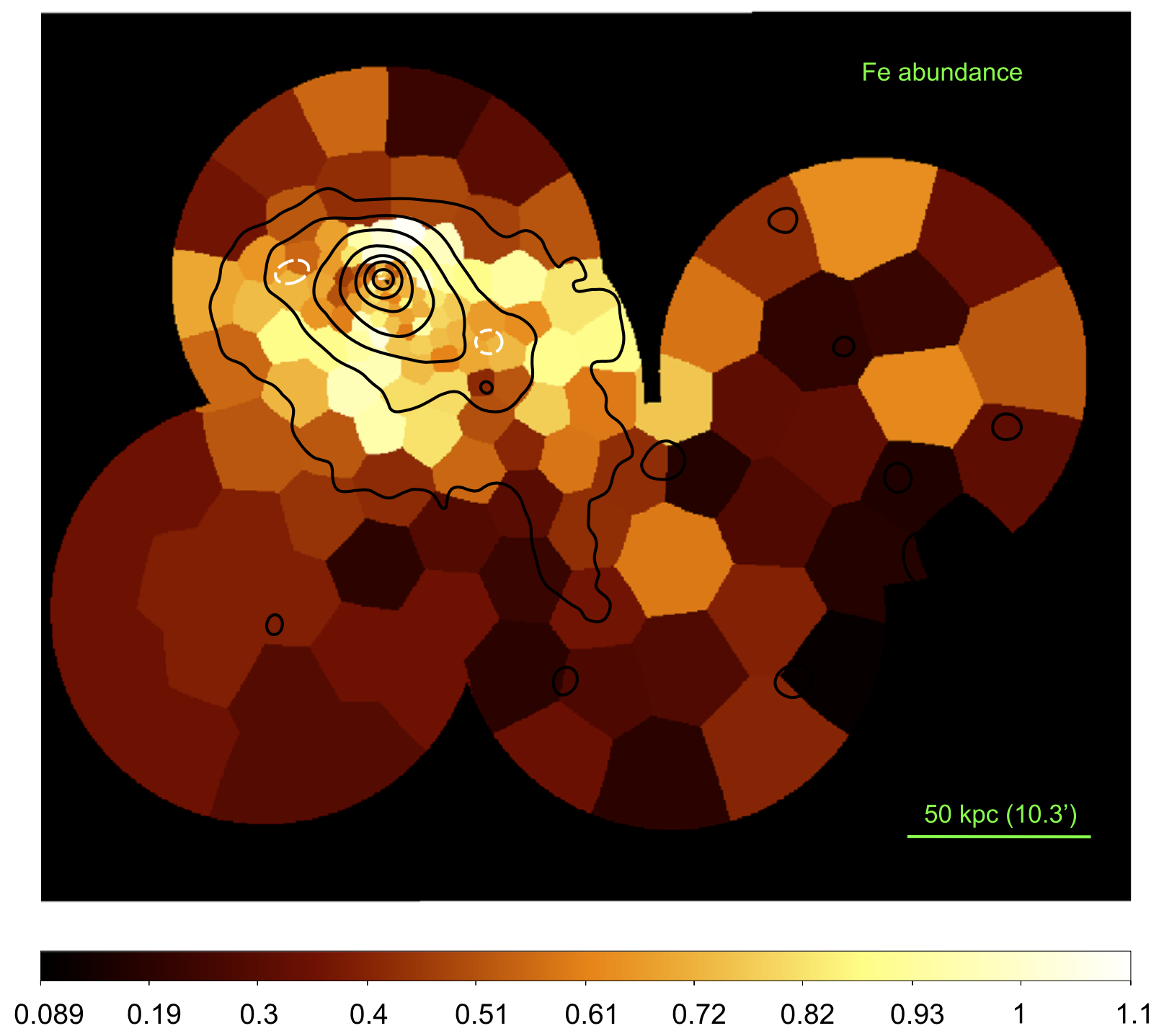}
     \caption{XMM-Newton spectroscopic map of the Fe abundance in M49 in units of the Solar abundance of \cite{Asplund:2009}, derived assuming a single temperature model. X-ray contours in the 0.7-–1.3 keV energy band are overlaid in black. The Fe distribution is elongated in the direction of the AGN ghost cavities (denoted by white dashed circles), with an additional extension towards the west/southwest on larger scales, likely related to a ram-pressure or slingshot tail as the galaxy is falling into the Virgo Cluster. Figure reproduced with permission from \cite{Su:2019}.}
     \label{fig:m49}
 \end{figure}
 
Of course, mergers and AGN feedback can work in unison. For instance, in M49, the 2D Fe abundance map derived by \citet{Su:2019} using XMM-Newton \citep[covering a significantly larger field than that in][ discussed above]{Gendron-Marsolais:2017} suggests both the presence of a metal enriched tail to the southwest, and a metal enhancement aligned with two outer ghost X-ray cavities along the NE-SW axis on smaller spatial scales (see Figure \ref{fig:m49}). The authors conclude that the tail gas can be traced back to the cooler and enriched gas uplifted from the BGG center by buoyant bubbles, implying that active galactic nucleus outbursts may have intensified the stripping process. On the other hand, \citet{Sheardown:2019} argue instead that M49 may host a slingshot rather than a ram-pressure tail. A similar case may be that of NGC507; again, no 2D metal abundance map is available, but \citet{Kraft:2004} report a gradient in the elemental abundance across a sharp arc-like X-ray surface brightness discontinuity with opening angle of 125 deg. Because that discontinuity is aligned with a low surface brightness radio lobe, the authors conclude that this `abundance front' can be explained by the transport of high-abundance material from the center of the galaxy due to the transonic inflation of the radio lobe; however, it was subsequently realized (e.g. see previous paragraph) that classical cold fronts are likely to produce such abundance arcs as well. The abundance feature in NGC507 is therefore not unusual and could be simply due to classical sloshing, or to an interaction between AGN feedback and past merging activity.

\subsection{Metal budgets} 
\label{s:mlr}

In the previous sections we reviewed the measurements of the abundances probing a fraction of the IGrM volume. As pointed out early by \citet{Arnaud:1992} the physically meaningful quantity for the study of the IGrM (and of the ICM) are the metal (iron or other chemical elements) mass and stellar mass present in the groups (and clusters). The ratio of the iron mass and stellar mass is directly linked to a fundamental quantity in chemical evolution models, the iron (or other chemical elements when measured) yield which is the ratio of the total iron mass released by stars to the total stellar mass formed for a given stellar population
\citep[see][and references therein]{Portinari:2004,Renzini:2014,Ghizzardi:2021}: 

\begin{equation}
    \mathcal{Y}_{\rm Fe} = \frac{M^{\rm star}_{\rm Fe,500} + M_{\rm Fe,500}}{M_{\rm star,500}(0)} ,
\end{equation}

where $M_{\rm Fe,500}$ is the iron mass enclosed within $r_{500}$ in the ICM/IGrM, $M^{\rm star}_{\rm Fe,500}$ is the iron mass locked into stars, $M_{\rm star,500}(0)$ is the mass of gas that went into stars whose present mass is reduced to $M_{\rm star,500}$ by the mass return from stellar mass loss, i.e. $M_{\rm star,500}(0) = r_o M_{\rm star,500}$, where $r_o$ is the  return factor. We take $r_o = 1/0.58 $ following \citet{Renzini:2014} and \citet{Maraston:2005}. A caveat should be made that the iron yield can be matched to a theoretical prediction only if we are able to make a full inventory under the assumption of a closed system. If iron can leave the system or just does not reside within the radius used to make the estimate we can not draw a conclusive inference. This is particularly the case at the scale of groups as we discuss further in this section.

One can then measure $M_{\rm Fe,500}$ either by multiplying a representative deprojected gas-mass weighted iron abundance times the total gas mass of the system within $r_{500}$ \citep{Renzini:2014} or by taking fully into account the radial dependence of the deprojected iron abundance and gas mass \citep{DeGrandi:2004,Ghizzardi:2021}:

\begin{equation}
     M_{\rm Fe} (<R) = 4\pi A_{\rm Fe} m_{\rm H} {Z_{\odot}}~ \int_0^R Z_{\rm depro}(r) ~n_{\rm H}(r)~ r^2 dr, 
\label{eq:mfe}
\end{equation}

where $Z_{\rm depro}$ is the deprojected abundance profile, $A_{\rm Fe}$ is the atomic weight of iron, and $m_{\rm H}$ is the atomic unit mass.
The hydrogen density $n_{\rm H}$ is derived from the gas density $n_{\rm gas}$ through the usual relation $ n_{\rm gas}= \left(1 + n_e/n_{\rm H} \right)n_{\rm H} = 2.21 n_{\rm H}$, where $n_e$ is the electron density; $n_{\rm gas}$ is obtained through deprojection. In the latter case $M_{\rm Fe,500}=M_{\rm Fe} (<r_{500})$.
For the measurement of $M^{\rm star}_{\rm Fe,500}$
it is usually assumed that the average iron abundance in clusters and groups stars is solar \citep[for the validity and limitations of this assumption see for example][and references therein]{Maoz:2010}. This iron abundance of the stars is then multiplied by the total stellar mass enclosed within $r_{500}$. The latter value can again be estimated through two approaches. The first one performs a flux measurement for each galaxy in a given optical band and calculates the mass of the galaxy through the Spectral Energy Distribution (SED) fitting, the mass of the galaxies are then summed together \citep{vanderBurg:2015,Ghizzardi:2021}. The second approach calculates the total luminosity in a given optical band by integrating the luminosity function of the red cluster galaxies, summing the contribution of the BCG and possibly of ICL and then multiply for an assumed stellar-mass-to-light ratio in the same optical band \citep[see for example][and referencese therein]{Renzini:2014}.
Both quantities are deprojected assuming a generalized or a simple Navarro-Frenk-White (NFW) distribution for the galaxy and optical light distribution.
These observational estimates can be compared with the expected theoretical estimate based on the current understanding of stellar nucleosynthesis. We take the values reported in \citet{Ghizzardi:2021} for the $\mathcal{Y}_{\rm Fe}$ based on the derivation by \citet{Renzini:2014} and \citet{Maoz:2017}. $\mathcal{Y}_{\rm Fe}$ is computed as the product of the Fe mass produced by a SN explosion, $y$, and the number of SN events produced per unit mass of gas turned into stars, $k$. 
Both contributions from Ia and CC SN are considered. Thus, $\mathcal{Y}_{\rm Fe}$ can be written as:

\begin{equation}
  \mathcal{Y}_{\rm Fe} = y_{\rm Ia}  \cdot  k_{\rm Ia} + y_{\rm CC}  \cdot k_{\rm CC},  
\label{eq:iron-yield}
\end{equation}
where Ia and CC subscripts refer to the two different SN types. 
For Ia we assume $y_{\rm Ia} = 0.7 \, M_\odot$ and $k_{\rm Ia} = 1.3\times 10^{-3} \,  M_\odot^{-1}$.  \citet{Renzini:2014},  see also \citet{Greggio:2011}, suggest a possible higher 
$k_{\rm Ia}$ value of $2.5\times 10^{-3} \, M_\odot^{-1}$. 
For CC SN we assume $ y_{\rm CC} = 0.074 \,  M_\odot $ and $ k_{\rm CC} = 1.0 \times 10^{-2}  \,  M_\odot^{-1}$ . Substituting the above values in Eqn. \ref{eq:iron-yield} and dividing by the solar abundance we get $\mathcal{Y}_{\rm Fe,\odot}= 0.93 \, Z_\odot$. 
An higher estimate can be obtained by assuming that SNIa rate is higher in clusters with respect to the field \citep{Maoz:2017,Friedmann:2018,Freundlich:2021}. If, following \citet{Freundlich:2021}, we assume a SNIa rate per unit mass of $k_{\rm Ia} = 3.1 \pm 1.1 \times 10^{-3} \,  M_\odot^{-1}$, we derive $\mathcal{Y}_{\rm Fe,\odot}= 2.34\pm0.62 \, Z_\odot$.  

 \begin{figure}[h]
 \centering
     \includegraphics[width=0.9\textwidth]{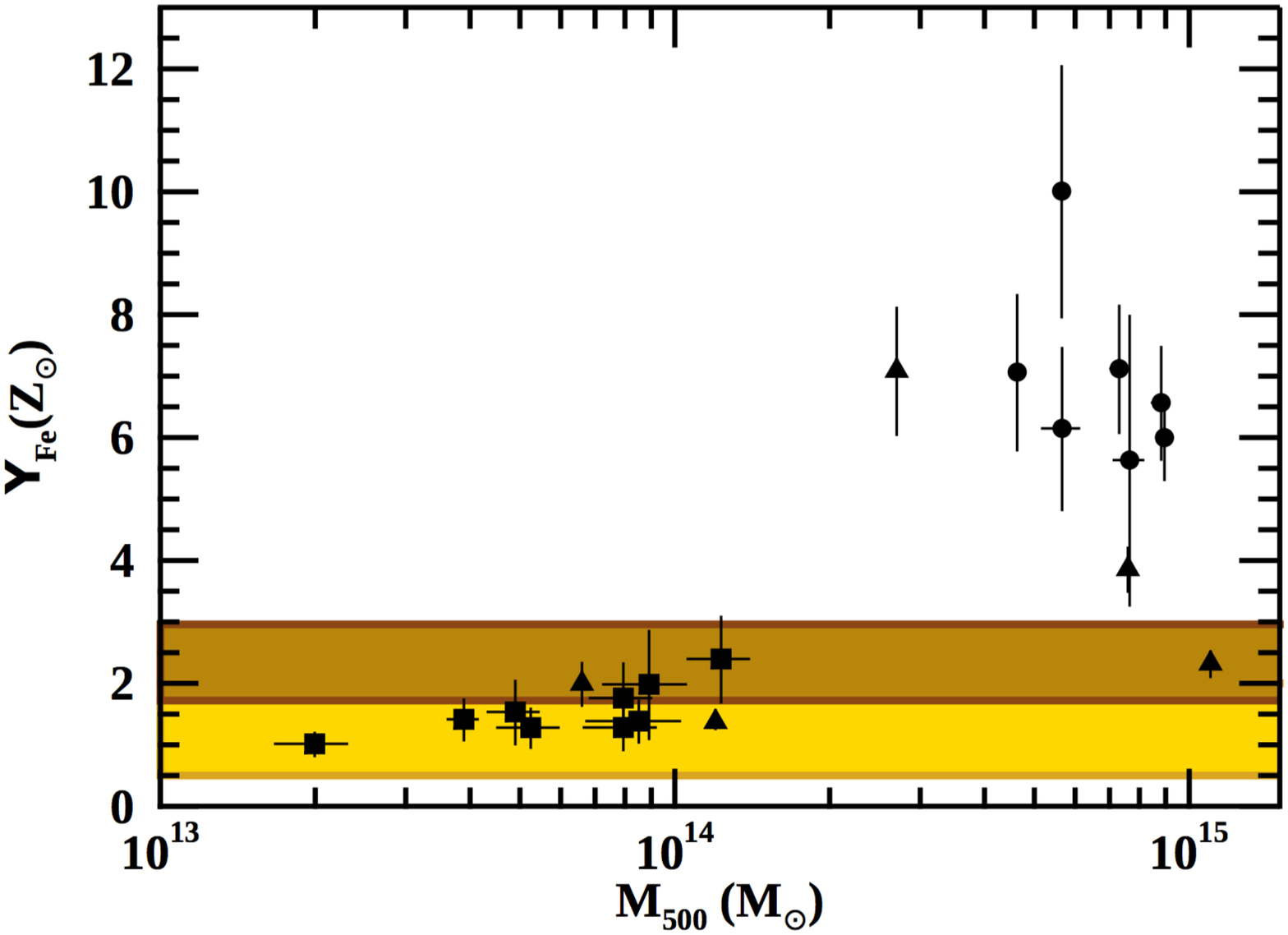}
     \caption{Effective iron yield $\mathcal{Y}_{\rm Fe,\odot}$ for the clusters in the sample of \citet{Ghizzardi:2021} (circles), for the groups in the sample of \citet{Renzini:2014} (squares) and for the additional objects (NGC 1550, MKW 4, Hydra A, Perseus and Coma in order of increasing mass) in the sample of \citet{Sasaki:2014} (triangles) as a function of the total mass of the system. We estimated stellar masses for the objects of \citet{Sasaki:2014} converting their $L_K$ optical luminosities to stellar masses, assuming a stellar mass-to-light ratio in the K-band of 1 consistent with stellar population models for a Kroupa IMF \citep{Maraston:2005} and with observational results \citep{Humphrey:2006b,Gastaldello:2007,Nagino:2009}.The yellow band shows the expected value computed through the SN yields derived from \citet{Maoz:2017} and \citet{Renzini:2014}; the brown band represents the expected value derived assuming a higher SNIa rate in galaxy clusters than in the field, following \citet{Freundlich:2021}.
}
     \label{fig:ironyield}
 \end{figure}

We report in Figure \ref{fig:ironyield} the effective iron yield for the sample of clusters studied in \citet{Ghizzardi:2021}, for the sample of groups studied in \citet{Renzini:2014}
with iron abundance measurements in the IGrM derived by \citet{Sun:2012} and for the objects in \citet{Sasaki:2014}. This plot seems to show apparently that for groups there is no discrepancy with the theoretical expectations as it is dramatically the case at the cluster scale.
The large amount of intra-cluster iron is difficult to reconcile with the metal production of stars seen in clusters today and it is posing a long standing challenge, with several unconventional solutions proposed such as a very different IMF in clusters or a significant contribution by pop III stars or pair-instability supernovae \citep[see the review by][and references therein]{Mernier:2018Review}. However these results should be treated with caution as not only the measurement of $\mathcal{Y}_{\rm Fe,\odot}$ is prone to systematic errors \citep[see the exaustive discussion in ][]{Ghizzardi:2021} but also the different trends of stellar and gas mass as a function of total mass play a fundamental role.
In particular the low baryonic fraction of groups within $r_{500}$ with respect to the cosmic baryon fraction does not allow to draw strong conclusions.
Clearly this plot should be more populated in particular at the mass scale of groups with robust measurements of the iron abundance consistently out to $r_{500}$. 

Indeed both the iron abundance and to a lesser extent the total iron mass in the IGrM do not depend only on the total amount of iron produced, but also on its dilution with pristine gas, its ejection due to non-gravitational feedback by AGN  and SN and the different phases in the gas.
All these effects should be taken into account 
 \citep[e.g.,]{Fabjan:2010}. 

Another related quantity exploits directly the luminosity of the system either in the optical (B) or infrared (K) bands: it is the ratio of the iron mass to the total light of the cluster/group \citep[IMLR,][]{Ciotti:1991,Renzini:1993}. If different metals in addition to iron are measured then the specific element MLR can be estimated, like for example O and Si MLRs.
These quantities are even more important to consider given the trends of stellar and gas fractions (and their sum, the baryon fraction) as a function of total mass which clearly mark the scale of groups as a crucial one in comparison with clusters. The derived baryon fraction for rich clusters is consistent with the cosmic baryon fraction, $\Omega_b/\Omega_M \sim 0.15$ \citep{Planck:2020} as obtained with X-ray, optical and infra-red observations \citep[e.g., ][see also the companion reviews by \citet{Lovisari:2021} and \citet{Eckert:2021}]{Vikhlinin:2006,Gonzalez:2007,Pratt:2009,Ichikawa:2013, Okabe:2014,Ghirardini:2019,Eckert:2019}
On the other hand, groups are characterized by higher stellar mass and lower IGrM mass fractions than rich clusters, and the number of baryon fraction tends to be lower with smaller groups \citep[e.g., ][]{Lin:2003,Giodini:2009,Dai:2010,Renzini:2014}.
Explanations of this discrepancy are suggested in the above references as follows: 1. different physical processes depending on the system mass, like for example a different efficiency of baryon-to-stars conversion; 2. observational data missing for fainter sources (as for example the intra-cluster light component) and for the IGRM at large radii; 3. systematic errors for the mass estimations; 4. non-gravitational heating and metal mixing by AGN feed back \citep[e.g., ][see \citet{Eckert:2021}]{McCarthy:2007,Fabjan:2010}, but a definitive solution has not been reached yet. As for the non-gravitational effect such as AGN feedback, the entropy profiles are good probe to estimate for each group and cluster, and we describe it in the following paragraph.

\begin{figure}[h]
 \centering
     \includegraphics[width=\textwidth]{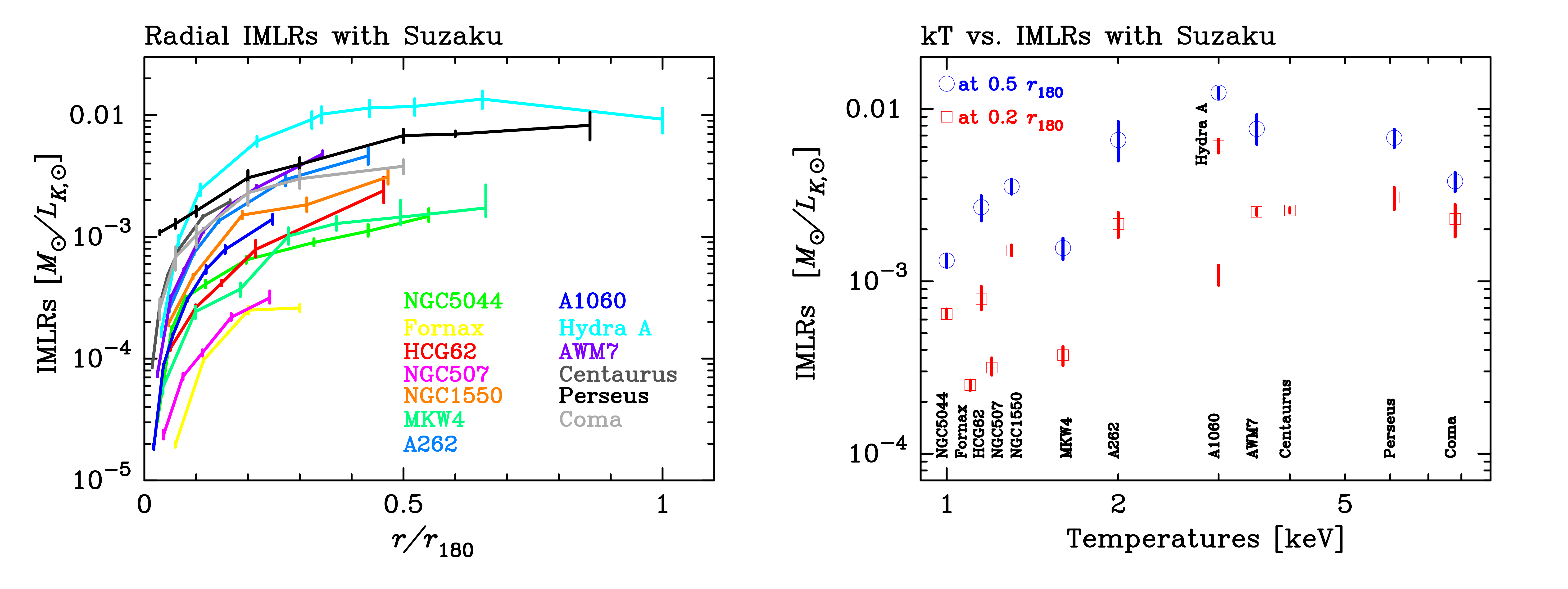}
     \caption{
(Left:) Radial IMLR profiles with Suzaku X-ray observations and K-band luminosity of the member galaxies from 2MASS data \citep[][]{Sato:2009, Sato:2010, Murakami:2011, Sakuma:2011, Matsushita:2013, Sasaki:2014}.
(Right:) Temperature dependence of the IMLRs for groups and clusters in 0.2 or 0.5~$r_{\rm 180}$ with Suzaku.}
     \label{fig:mlrs}
 \end{figure}

Historically, \citet{Arnaud:1992} found that the total iron mass in the ICM is proportional to the total luminosity of the early type galaxies in the clusters. And the IMLR in the B-band had larger values in clusters than in groups, mainly caused by the biased low early global abundance measurements (see \S\ref{s:obs_hist}) at the group scale \citep[e.g., ][]{Renzini:1997,Makishima:2001}. 
The current state of the art of measurements of the IMLRs is achieved by combining near-infrared (K-band) luminosities \citep[more directly related to the bulk of the stellar mass in early type galaxies,][]{Lin:2004,Nagino:2009} obtained with the two micron all sky survey (2MASS) and the measurements performed by Suzaku extending to the outer regions of nearby clusters and groups \citep[e.g., ][]{Sato:2009, Sato:2010, Matsushita:2013}. Figure \ref{fig:mlrs} reports the IMLRs thus obtained: there is a general increase with radius and poorer clusters and groups also have lower IMLRs within the 0.2 $r_{180}$ region. On the other hand, the IMLRs of groups and clusters in the outer region at $r>0.5~r_{180}$ seem to be closer to each other. Figure \ref{fig:mlrs} suggests that poorer systems (groups and clusters with fewer member galaxies) could not hold the gas including metals due to the relatively shallower potential in their assembly process. In a following work \citet{Sasaki:2014} showed that the same systems have lower IMLRs and larger entropy excess, which is a signature of the non-gravitational energy input in the central regions of groups during the assembly stage.

Abundance ratios of silicon to iron in clusters and groups look similar to each other in $r<r{500}$ 
(see \S\ref{s:composition}), and the silicon mass in poorer systems has a lower value than those in larger systems (the same trend as for the iron mass, caused by the lower gas mass). Because significant fractions of oxygen and silicon are mainly synthesized by SNcc, they are good indicators to estimate the amount of massive stars in the past. \citet{Renzini:2005} calculated the oxygen and silicon MLRs under the assumption of the Salpeter initial mass function with the slope of the power-law shape to be $-2.35$. If we assumed the silicon to iron abundance ratios to be $\sim 1$ up to the virial radius \citep[e.g., ][]{Matsushita:2013,Mernier:2017}, the expected silicon mass to light ratios (SiMLR) in rich clusters agree with the estimate derived by \citet{Renzini:2005}, as shown in \citet{Matsushita:2013} and Sasaki et al. (2021). However for groups neither iron nor silicon have been observed yet out to and beyond $r_{500}$ with the exception of a handful of systems. To study the metal enrichment history of the ICM and IGrM, we need to measure oxygen profile, mainly produced by SNcc, to the virial radius and beyond since the whole clusters and group include the metals synthesized in the cluster and group formation phase. 
To summarize, in order to progress in the understanding of the chemical enrichment of both groups and clusters, we need to measure the gas and metal MLRs to the virial radius for both clusters and groups to high redshift of $z\sim 2$ with the next generation of X-ray instruments (see \S\ref{s:tele}).

\subsection{High resolution spectroscopy: Current observational results with RGS} 

Undoubtedly, the need for high spectral resolution (in particular the capability to resolve the Fe-L complex) is absolutely crucial to provide accurate constraints on chemical abundances (and the physics of the enrichment) in groups and elliptical galaxies. While waiting for the exploitation of micro-calorimeters onboard \XRISM\ and \Athena, one should keep in mind the valuable potential of the Reflection Grating Spectrometer (RGS) instrument onboard \XMM\ to deliver high resolution spectra. Formally, the RGS has a spectral resolution of $\sim$3~eV. In the case of extended sources, however, the slit-less characteristic of this instrument induces a convolution of a given line profile with the spatial distribution of its surface brightness along the dispersion direction of the detector. Although this makes the RGS abundance measurements of extended clusters rather challenging \citep[yet still feasible; e.g.][]{Simionescu:2019}, groups are by definition more compact and are less affected by this instrumental broadening. The spectral window of the instrument (typically $\sim$6--30 $\mathring{A}$, corresponding to $\sim$ 0.4--2 keV) is both an advantage -- as it covers the O VIII (and O VII), N VII, and even C VI lines which are difficult or impossible to detect with CCD instrumental responses -- and a drawback -- as the continuum is challenging to constrain within this band. Consequently, the power of RGS resides less in the measurements of the IGrM absolute abundances than in the measurements of their (relative) N/Fe, O/Fe, Ne/Fe, and Mg/Fe ratios. Although discrepancies of absolute Fe measurements between RGS and CCD-like instruments may in principle lead to discrepancies in their respective X/Fe ratios, we note a generally good agreement in the latter case \citep[e.g.][]{Mernier:2016a}.

Abundances measured using RGS have been for instance reported on individual poor systems \citep{Xu:2002,Werner:2009,Grange:2011} as well as in larger samples \citep{Tamura:2003,Sanders:2010,Sanders:2011}. The CHEERS sample, which was constructed specifically to ensure a $>$5$\sigma$ detection of the O VIII line in each system, provided interesting measurements in this respect. \citet{Mao:2019} obtained reliable constraints ($>$3$\sigma$) on the N/Fe ratio in six galaxies (M\,49, NGC\,4636, NGC\,4649) and groups (NGC\,5044, NGC\,5813, and NGC\,5846), as well as in M\,87 and one cluster (A\,3526). Unlike all the other ratios known in the IGrM which are typically close to Solar (Sect. \ref{s:composition} and Fig. \ref{fig_CHEERS}), the average N/Fe is clearly super-Solar. This strongly suggests that the bulk of nitrogen originates from an enrichment channel that is separate from the usual SNcc and SNIa contributions -- very likely AGB stars. The O/Fe ratio was investigated (in groups and ellipticals, but also in more massive systems) by \citet{dePlaa:2017}, and was found to be consistent with Solar, thus in line with the picture of the hot gas chemical composition remaining uniform with mass (Sect. \ref{s:composition}), despite a significant scatter from system to system.

 \begin{figure}[h]
 \centering
     \includegraphics[width=0.85\textwidth]{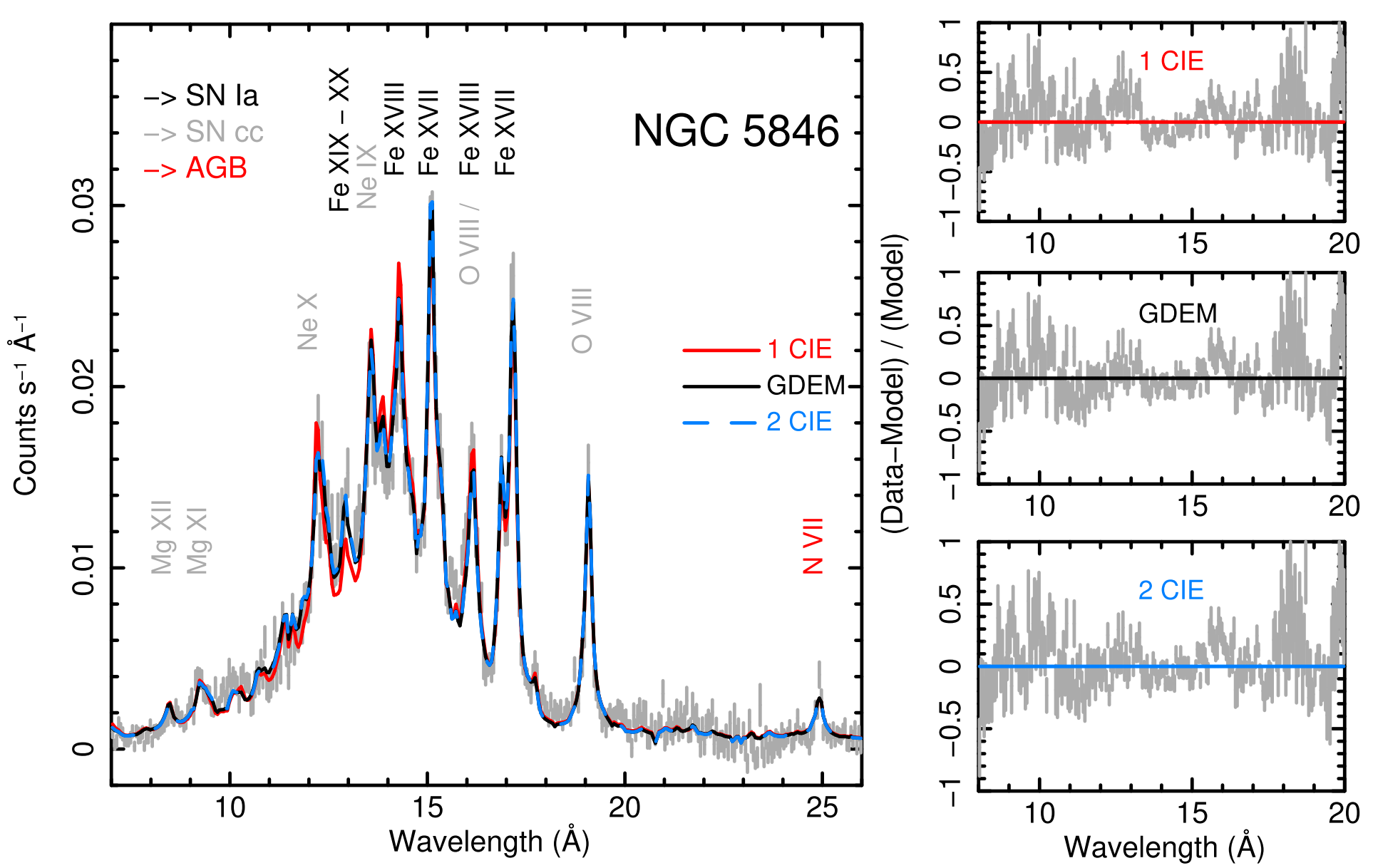}
     \caption{An example of RGS spectrum from a deep observation of NGC\,5846, with indication of its main emission lines (and the stellar origin of their corresponding element). Spectra from the RGS\,1 and RGS\,2 instruments were combined before successive fits assuming models with one temperature (1\,CIE), two temperatures (2\,CIE), and a Gaussian-like multi-temperature structure (GDEM). Residuals are shown in the right panels. Figure reproduced with permission from \citet{dePlaa:2017}. }
     \label{fig_RGS}
 \end{figure}

Despite the valuable ability of RGS to measure accurately important ratios (in particular N/Fe and O/Fe, see Figure \ref{fig_RGS}), its limited spectral window -- coupled to its sensitivity to the spatial extent of the source and the difficulty to perform spatial spectroscopy \citep[see however][]{Werner:2006b,Ahoranta:2016} -- makes this instrument taken at its best advantage when combined to CCD measurements. Nevertheless, RGS offers the unique advantage to reveal a glimpse of the main transitions populating the Fe-L complex at groups (and clusters) temperature regime(s). This is particularly essential, not only to refine our science prediction expected with micro-calorimeter instruments like Resolve onboard \XRISM\ (Sect. \ref{s:XRISM}), but also to pursue the improvement of our spectral models in this crucial spectral band even before the release of \XRISM, particularly by comparing updated atomic calculations with (i) laboratory measurements and (ii) state-of-the-art observational data \citep{Gu:2019,Gu:2020}.




\section{Theoretical framework and simulations} \label{s:theory} 
\subsection{High-resolution hydrodynamical simulations and small-scale astrophysics} 
\label{s:HDsims}
While \S\ref{s:cosmosims} focuses on the large-scale metal origins and evolution from the high-$z$ universe via cosmological particle simulations, here we discuss the metal abundance evolution in terms of small-scale (astro)physics and tracer advection by means of high-resolution (HR) hydrodynamical (HD) grid simulations in single halos. Resolutions in such simulations can reach down to the pc scale and often employ adaptive/static mesh refinement, together with finite-volume Godunov methods with third-order (or higher) accuracy. 
Given the high resolution (and small timesteps), such simulations are mainly focusing on the cores of galaxy groups ($r < 0.1\,\r500$) and are well suited to track the turbulence mixing, shocks, entrainment, and detailed feedback/feeding imprints such as AGN cavities and thermal instabilities.
We note that, in this \S\ref{s:HDsims}, we take a more physically-oriented approach, rather than following an historical sequence.

Before tackling typical HD simulation results, it is worth to review a few common physical and numerical properties of metals leveraged by investigations of different groups. A key small-scale feature of metals is that they are passive tracers of the hydrodynamical evolution; thus, they can be used akin to dyes/pollutants in fluid dynamics studies \citep[e.g.][]{Scheeler:2017}. 
In hydrodynamics \citep[e.g.][]{Warhaft:2000},
the equation describing the temporal rate of change of the metal tracer density is given by
(in the Eulerian/grid framework):
\begin{equation} \label{e:passive}
    \frac{\partial \rho_{Z}}{\partial t} + \div{(\rho_{Z} \boldsymbol{v}) = S_{Z}},
\end{equation}
where $S_{Z}$ is a general metal abundance source term; in Eq.~\ref{e:abu} and related paragraph below, we will unpack such a source term, which, in the IGrM, is mainly shaped by stellar feedback processes such as supernovae and stellar winds. 
In fluid dynamics, Eq.~\ref{e:passive} is also known as a conservation equation. In the weak compressibility case, the metals are purely advected along the Lagrangian stream, reducing Eq.~\ref{e:passive} to
\begin{equation} \label{e:Lagr}
\frac{\dd \rho_Z}{\dd t} \equiv \frac{\partial \rho_{Z}}{\partial t} + \boldsymbol{v} \cdot \boldsymbol{\nabla} \rho_Z = S_Z.
\end{equation}
We note that in localized IGrM regions with $\div{\boldsymbol{v}} \ne 0$, the pollutants may also experience compressions or rarefactions, hence tracing not only smooth bulk processes (subsonic turbulent eddies) but also nonlinear in-situ features (shocks and cold fronts). 
In the HD grid simulations, such iron density is implemented via a normalized scalar $Z \in [0,1]$ tied to each cell gas density such as $\rho_Z = Z\,\rho$ \citep[see][]{Gaspari:2011a}. Ought to high-order Godunov schemes such as the Piecewise-Parabolic Method (PPM; \citet{Colella:1984}) numerical diffusion is kept at low levels compared with physical diffusion, e.g., due to turbulence.
Further, despite the large diversity of elements, IGrM HR numerical studies often use the approximation $Z \approx Z_{\rm Fe}$, since iron has one of the strongest line emissions among heavy elements -- especially in hot plasma halos -- which can be robustly constrained via X-ray spectra (\S\ref{s:obs}).
We note that, while metals are a dynamical passive tracer, they contribute significantly to the line cooling of the gas below $T<1\ {\rm keV}$ \citep[e.g.][]{Sutherland:1993}, hence accelerating the IGrM condensation cascade.

Typical HR HD simulations focus on the group core, where the central galaxy contributes to a substantial amount of metals later dispersed in the diffuse IGrM, which are mainly produced via supernovae (SN) explosions and winds from red giant stars (SWs). An exemplary and well modeled system is the nearby galaxy group NGC 5044, with the homonymous central galaxy dominating over the many smaller satellites.
In several ETGs/BGGs analytic studies and non-cosmological HD simulations [e.g.~\citealt{Loewenstein:1991}, \citealt{Ciotti:1991}, \citealt{Renzini:1993,Brighenti:1999a,Brighenti:2002,Mathews:2003,Gaspari:2011b,Gaspari:2012b,Pellegrini:2020}], a common implementation of Eq.~\ref{e:Lagr} is to recast it in terms of the astrophysical IGrM abundance of the $i$th-element (by mass in Solar unit):
\begin{equation} \label{e:abu}
    \frac{\dd Z_i}{\dd t} = (\underbrace{N_\ast\,\alpha_\ast}_{\rm stellar\ winds} + \underbrace{N_{\rm SN}\,\alpha_{\rm SN}}_{\rm supernovae})\,\frac{\rho_\ast}{\rho},
\end{equation}
where the two normalization factors related to stellar winds and supernovae are respectively $N_\ast=Z_{\ast,i} - Z_i$ (with $Z_{\ast,i}$ the stellar abundance) and $N_{\rm SN}=y_{{\rm SN},i}/(Z_{i,\odot} M_{\rm SN})$ (with $y_{{\rm SN},i}$ the SN yield in $\msun$ and $M_{\rm SN}$ the ejected supernova mass).
The BGG stellar density $\rho_\ast$ is usually modeled via a de\,Vaucoulers profile \citep{Mellier:1987}.
The specific injection rates due to SWs and SN are 
$\alpha_\ast \approx 4.7\times10^{-20}(t/t_{\rm now})^{-1.3}$\,s$^{-1}$ and
$\alpha_{\rm SN} \simeq 3.2\times10^{-20}\,r_{\rm SN}(t) (M_{\rm SN}/\msun)\,\Upsilon_B^{-1}\ {\rm s^{-1}}$, where $\Upsilon_B^{-1}$ is the optical stellar mass-to-light ratio in the B band \citep[e.g.][]{Mathews:2003,Greggio:2005,Mannucci:2005,Humphrey:2006a}, respectively.
As introduced above, iron is one of the best metals to leverage as dynamical tracer in HR simulations: in BGGs/ETGs (which have star-formation histories peaked at early times) SNII are mostly consumed at high redshifts, while SNIa  -- exploding in binary systems with a white dwarf -- drive the BGG long-term iron enrichment. The local SNIa rate is $r_{\rm SNIa} \sim 0.1 (t/t_{\rm now})^{-1.1}$\,SNU (per 100 year and stellar luminosity $10^{10}\,L_{\rm B,\odot}$; e.g.~\citet{Greggio:2005,Mannucci:2005,Humphrey:2006a}). The iron normalization values for the SNIa ($M_{\rm SNIa} = 1.4\,\msun$) are $y_{\rm SNIa, Fe} = 0.7\,\msun\sim10\,y_{\rm SNII, Fe}$ and $Z_{{\rm Fe},\odot} = 1.83\times10^{-3}$.

 \begin{figure}[!ht]
     \subfigure{\hspace{-1.0cm}
     \includegraphics[width=1.12\textwidth]{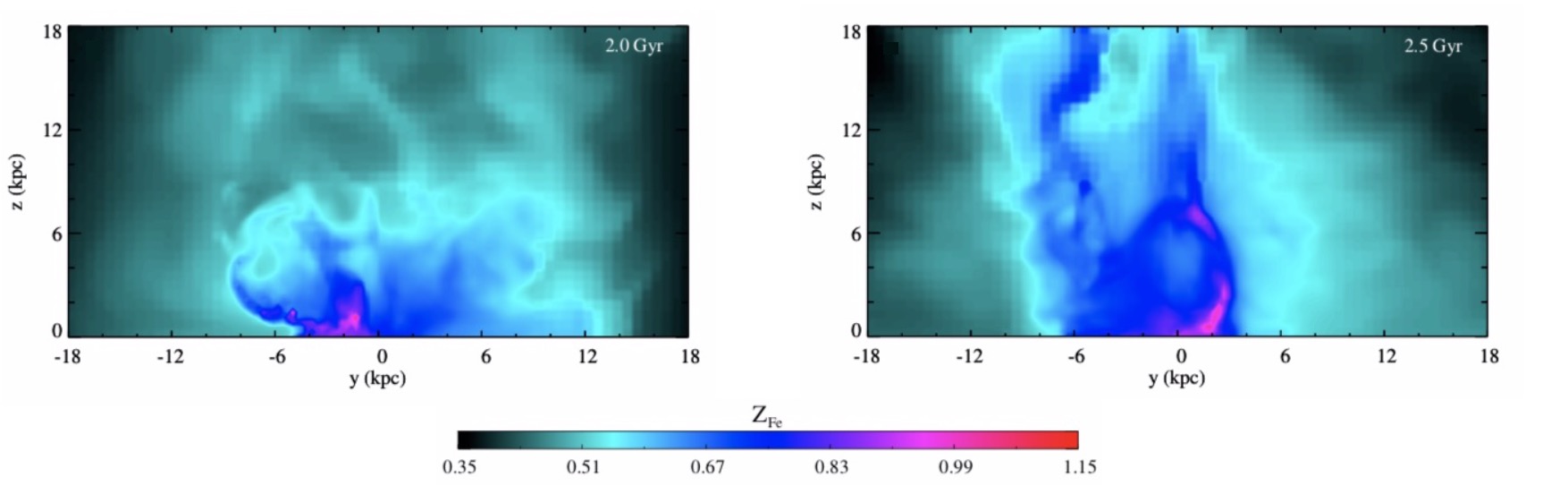}}
     \caption{Emission-weighted iron abundance during a typical HR HD 3D simulation of self-regulated AGN feedback in a median 1\,keV galaxy group akin to NGC\,5044 (adapted from \citet{Gaspari:2012b}), showing two typical stages. The black regions denote the diffuse, primordial iron background ($Z_{\rm Fe} < 0.3\,Z_\odot$).
     Left: turbulence-driven period, during which mixing dominates, gradually washing out the anisotropic features and restoring azimuthal symmetry.
     Right: the meso-scale (sub-kpc) AGN outflows have inflated a common X-ray cavity in the IGrM, generating a thin metal-rich rim (coincident with the compressive cocoon shock) and anisotropic iron uplift from the BGG outwards in the extended IGrM.
     }
     \label{f:zout}
 \end{figure}

We now review the results of HD simulations leveraging the metal tracing framework. 
In HR HD simulations, metals are a crucial tool to unveil and understand the kinematics of major physical processes, such as AGN feedback, SMBH feeding, and IGrM turbulence. While the detailed AGN self-regulation thermodynamics (cooling and heating cycle) is tackled in the companion \citet{Eckert:2021} review (see also the unification diagram in \citet{Gaspari:2020}), here, we focus on the main IGrM cores kinematics features.
The SMBH/AGN at the center of each BGG or ETG is fed recurrently via chaotic cold accretion \citep[CCA;][]{Gaspari:2013_cca}, i.e., the filamentary/cloudy condensation rain that is generated via \textit{nonlinear} turbulent thermal instability in the IGrM hot halo (\citet{Gaspari:2017_cca,Voit:2018}; see also \citet{McCourt:2012,Sharma:2012} for linear thermal instability simulations). Such frequent CCA clouds trigger the AGN feedback response by re-ejecting substantial amount of mass and  energy via ultrafast outflows and relativistic jets \citep[e.g.][]{Tombesi:2013,Sadowski:2017}. At the macro scale of tens kpc, such entrained outflows/jets use their mechanical ram pressure to generate a diverse range of astrophysical phenomena (buoyant X-ray cavities, weak transonic shocks, turbulent eddies), which recurrently re-heat the IGrM halo and quench cooling flows/star formation throughout the several Gyr evolution [e.g.~\citealt{Churazov:2001,Bruggen:2003}, \citealt{Brighenti:2006}, \citealt{McNamara:2007,McNamara:2012,Gitti:2012,Gaspari:2012a,Barai:2016,Yang:2019}].

The macro-scale AGN feedback deposition channels are difficult to disentangle through simple temperature or surface brightness maps. The metal tracers are instead able to unveil in a clear manner such feedback features. Indeed, while the BGG produces a continuous reservoir of metals/iron in the core of the galaxy group, the self-regulated AGN outflows uplift them (Eq.~\ref{e:passive}) outwards, on top of the low-$Z$ background, thus creating key contrast patterns and imprints.
Fig.~\ref{f:zout} exemplifies this during a typical HR HD simulation \citep{Gaspari:2012b} of AGN feedback -- self-regulated via CCA - in a galaxy group akin to NGC\,5044 (with a central BGG $M_\ast \simeq 3.4\times10^{11}\,\msun$).
In the right panel, the meso AGN outflows have just uplifted the iron generated in the core of the BGG (magenta) up to several tens kpc. The pattern is highly anisotropic and the enhancement can reach up to $\sim2\times$ values compared with the pristine background ($<0.3\,Z_\odot$, e.g.~\citet{Ghizzardi:2021}). Inhomogeneities are also  visible, particularly the thin metal-rich rim that envelopes the inflated buoyant bubble.
At variance, in the left panel, as the AGN outflows subside and CCA feeding is quenched via the previous AGN outburst, the cascading subsonic turbulence drives mixing, eventually washing out the inhomogeneities (cavity, cocoon, jet channel) and restoring the azimuthally symmetric IGrM halo `weather'. Subsequently, this enables another phase of gradual IGrM precipitation and hence boosted accretion, with the triggering of another AGN feedback cycle via the condensed material.
Both the above-shown anisotropic metal outflows and turbulent distributions have been found by a wide range of hot halo observations involving mechanical AGN feedback (\S\ref{s:maps}) and a diverse range of HD numerical studies and groups \citep[][]{Gaspari:2011a,Barai:2014,Taylor:2015,Valentini:2015,Duan:2018,Wittor:2020,Choi:2020}. We note that galactic SN-driven outflows, albeit weaker and difficult to spatially detect, can enhance the anisotropic enrichment, especially in low-mass halos \citep[e.g.][]{Melioli:2015,Husemann:2019}.

\begin{figure}[!ht]
     \centering
     \subfigure{
     \includegraphics[width=0.855\textwidth]{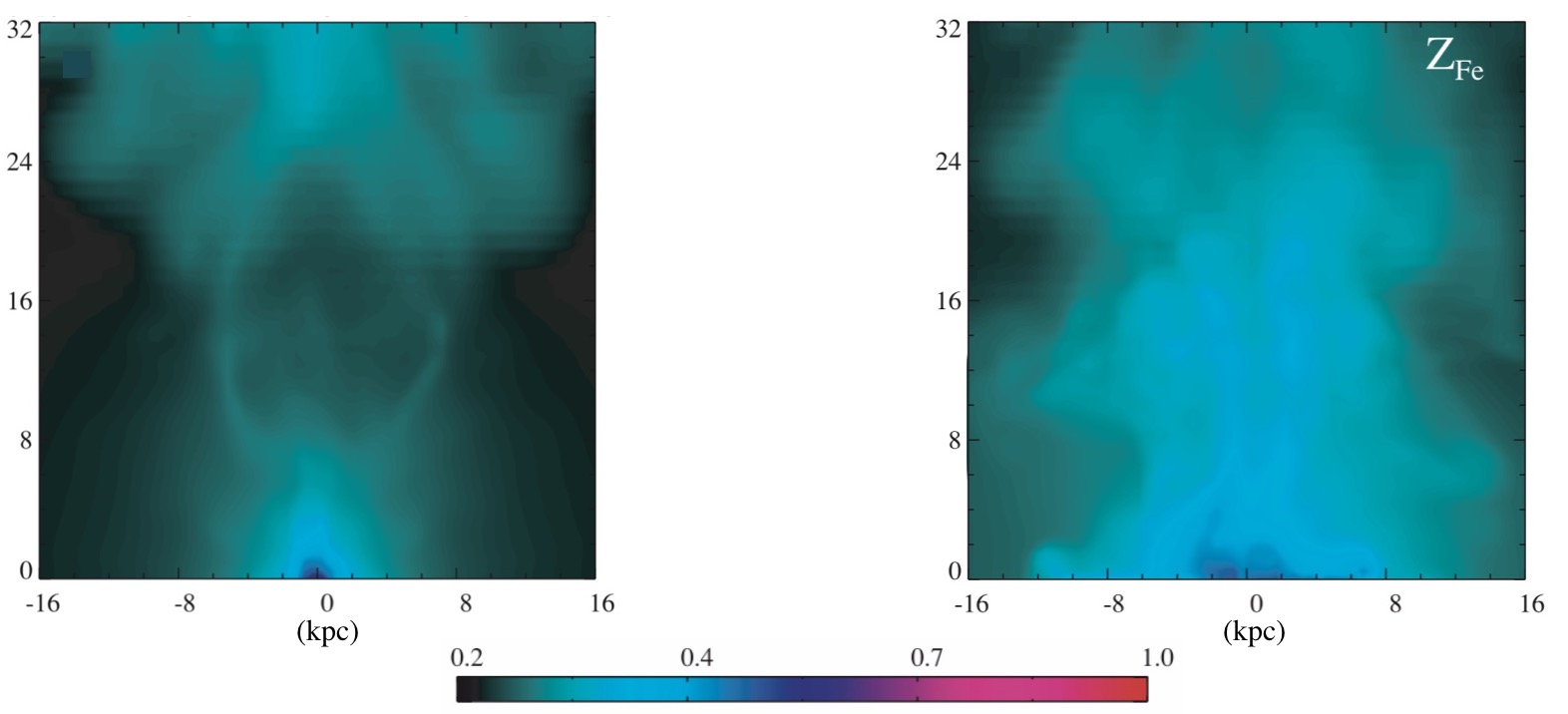}}
     \caption{The iron abundance projected maps can differentiate between different models of AGN feeding triggering and self-regulation, here in a common HR HD simulation of a massive galaxy group (adapted from \citealt{Gaspari:2011b}). Left: CCA feeding mode, driving intermittent and frequent AGN features such as cavities with metal-rich rims and trailing filaments. Right: Bondi feeding mode, driving a perennial, wide monolithic cylinder of metals into the group core (with no bubbles or cocoons).
     }
     \label{f:zfeed}
\end{figure}

Metal maps not only give us constraints on the effects of AGN feedback, but can also constrain different AGN feeding modes. Fig.~\ref{f:zfeed} shows two HR HD simulations testing two different models of AGN self-regulation and feeding in a massive galaxy group with extended IGrM ($L_{\rm x}\sim10^{43}\,\es$). 
In the left panel, CCA feeding/cold mode self-regulation drives a very intermittent duty cycle tightly related to the recent cooling rate in the group core (typically with pink-noise time power spectrum; \citet{Gaspari:2017_cca}). The rapid flickering of the AGN enables the formation of characteristic AGN feedback imprints, such as buoyant bubbles and cocoon shocks that are encased by metal-rich rims (cyan) protruding on top of the low abundance background. In addition, the buoyant bubble can dredge out trailing filaments of metals via the hydrodynamical \citet{Darwin:1953} effect ($V_{\rm trail} \sim 0.5\,V_{\rm b}$, where $V_{\rm b}$ is the bubble volume).
Conversely, hot-mode feeding (e.g., Bondi or ADAF; \citet{Bondi:1952,Narayan:2011}) drives a continuous AGN feedback evolution with a perennial, monolithic, wide and uniform cylinder of metals, without any signs of cavities or shocks. As for the thermodynamics \citep[cf.~the companion review][]{Eckert:2021}, observational constraints favor the former intermittent bubble duty cycle (via cold mode/CCA), rather than the latter quasi-continuous triggering mode (Bondi or hot-mode feeding) -- \citet{Simionescu:2008,Hlavacek-Larrondo:2011}, \citet{Gitti:2012,McNamara:2012,Liu:2019,Gaspari:2019}.

As introduced above, a key component of the metal circulation is turbulence, either generated by the AGN feedback or by the large-scale cosmological evolution (\S\ref{s:cosmosims}), which is worth to further dissect. Remarkably, turbulent motions generate two (seemingly) contrasting effects, but on different scales. On the one hand, turbulent motions promote the diffusion of metals, tending to equalize the radial abundance profile from a negative to null gradient \citep[e.g.][]{Rebusco:2005,Rebusco:2006}. On the other hand, turbulence induces local chaotic relative density fluctuations $\delta \rho/\rho$. As per above Eq.~\ref{e:Lagr}, metals can be considered on average akin to passive tracers of the HD density field, thus $\delta \rho_Z/\rho_Z\sim \delta \rho/\rho$. As shown by numerical and analytic studies \citep[e.g.][]{Gaspari:2013_coma,Zhuravleva:2014}, such stratified hot-halo fluctuations are linearly tied to the turbulent Mach number ${\rm Ma_t}$, hence the relative metal abundance can help us to constrain the level of turbulence in the IGrM too, $\delta \rho_Z/\rho_Z \propto {\rm Ma_t}$, with the slope of the Fourier spectrum constraining plasma processes such as thermal conduction \citep[][]{Gaspari:2014_coma2}. In the IGrM, the inferred 3D sonic Mach number of turbulence is ${\rm Ma_t} \sim 0.3-0.5$ \citep{Hofmann:2016,Ogorzalek:2017}, i.e., $\sigma_v$ of a few 100\,$\kms$.
This can complement upcoming spectral X-ray IFU studies carried out via \XRISM\ and \Athena\ (see \S\ref{s:tele}), with detailed synthetic observations already highlighting unprecedented features of metals and turbulence in hot halos \citep{Lau:2017,Cucchetti:2018,Roncarelli:2018,Mernier:2020b}. Moreover, constraining the turbulent metal evolution in the IGrM plasma phase enables to assess the kinematics of the top-down multiphase rain, since the condensed warm (H$\alpha$+[NII]) filaments and cold (CO, HI) clouds share analogous ensemble velocity dispersion \citep{Gaspari:2018,Tremblay:2018,Rose:2019,Simionescu:2019rev}.

While the large-scale cosmological evolution is discussed in the upcoming \S\ref{s:cosmosims}, it is worth to note here that at $r\gta100$\,kpc (and with Gyr frequency), the infalling sub-structures and interacting galaxies (in particular dry dark matter halos; see the HD simulation review by \citet{ZuHone:2016}) can induce significant amount of sloshing in the IGrM, hence creating large-scale metal anisotropies and tails that are often correlated with cold fronts/contact discontinuities or ram-pressure stripping features \citep{Ettori:2013_metals,Gastaldello:2013,Ghizzardi:2014,O'Sullivan:2014,DeGrandi:2016,Eckert:2017_ramP,Clavico:2019,Tumer:2020}. For the observational insights on related metallicity maps we refer the interested reader to \S\ref{s:maps}.

%

\subsection{Cosmological simulations and large-scale evolution}\label{s:cosmosims}

The metal content of cosmic structures has been addressed via complex large-scale cosmological hydrodynamical simulations as well~\cite[][]{Biffi:2018R}.
Cosmological simulations~\cite[][]{Dolag:2008} allow us to study and predict the formation and evolution of galaxies and galaxy systems, such as groups and clusters, within the large-scale cosmological framework~\cite[][]{Borgani:2011,Vogelsberger:2020}, while consistently accounting for a large variety of physical processes shaping the baryonic matter component
--- from gas cooling, to star-formation and BH evolution, to energy feedback.
In particular, chemical evolution models are needed to consistently follow the production and evolution of the metal content in the stellar and gaseous components, which has important consequences on the cooling properties of the gas, on the conversion of gas into stars, and therefore is linked to the thermo-dynamical structure of galaxy systems. 
Given the large dynamical ranges spanned in simulations of cosmological volumes, chemical evolution is typically treated via a sub-resolution model, similarly to other small-scale physical processes (like star formation or energy feedback).

Chemical evolution models have been introduced in cosmological simulations starting from the 90's~\cite[][]{Steinmetz:1994,Mosconi:2001}.
While early studies including chemical enrichment mainly focused on galaxies~\cite[][]{Steinmetz:1994,Kawata:2003}, soon Smoothed-Particle Hydrodynamics (SPH) simulations of 
large-scale structure and galaxy clusters started to include chemical evolution models as well~\cite[][]{Lia:2002,Valdarnini:2003,Tornatore:2004}. 
Despite different level of complexity, already in the early implementations, the metal production associated to both SNIa and SNcc was included, and the metal content of the gas was taken into account in the cooling process~\cite[e.g.][]{Scannapieco:2005,Oppenheimer:2006}. 
In~\cite{Tornatore:2004}, the contribution to chemical enrichment due to low- and intermediate-mass stars undergoing the AGB phase, as well as the treatment of metallicity-dependent stellar yields and mass-dependent stellar life-times were also included~\cite[][]{Tornatore:2007}.
Starting from the initial models that focused on total metallicity or iron abundance, an increasing level of detail has been reached over the years, with modern simulations typically following the production and evolution of several metal species separately (e.g. oxygen, silicon, etc.). 
Chemical enrichment models are typically based on three fundamental pillars: the initial mass function (IMF)~\cite[e.g.][]{Salpeter:1955,Kroupa:2001,Chabrier:2003}, and mass-dependent stellar life-times~\cite[e.g.][]{Tinsley:1979,Padovani:1993,Portinari:1998,Romano:2005_I} and metal yields~\cite[e.g.][]{Romano:2010_II,Nomoto:2013,Karakas:2014,Nomoto:2009,Nomoto:2018}.
In SPH simulations~\cite[][]{Springel:2010}, in particular, chemical evolution models are coupled directly with the star formation model, where every stellar particle is representative of a simple stellar population (SSP), that is a population of stars all characterized by the same age and metallicity. The assumptions on the IMF and on the stellar yields are required to predict the amount of metals generated by each SSP and the time-scale on which different enrichment channels (primarily SNIa, SNcc and AGB stars) release the metal mass into the surrounding gas elements~\cite[e.g.][for more details on the principal equations that describe the stellar evolution and metal production]{Tinsley:1980,Matteucci:2003}.

Results from cosmological hydrodynamical simulations on the chemical enrichment of cosmic structures, from galaxies to groups and clusters, can be very sensitive to the specific assumptions on the IMF or of stellar yields. 
In particular, these can affect the normalization of metallicity profiles and the value of global abundances. 
Changes in the underlying IMF functional form, affect directly the final ICM metallicity and abundance ratio profiles, for instance, due to different relative amounts of low- and high-mass stars~\cite[][]{Romano:2005_I,Romeo:2006,Tornatore:2007}.
The yield tables are also an important source of uncertainty~\cite[][]{Romano:2010_II} in the predicted integrated level of enrichment, as well as the supernova rates~\cite[see a recent investigation based on the Illustris Simulations by][]{Vogelsberger:2018}.
More importantly, the complex interplay with other gas thermal and dynamical processes treated in the simulations, such as energetic feedback or merging processes,
can substantially impact the spatial distribution of metals and therefore the gradients of the radial profiles, as further discussed in \S\ref{s:cosmo-res} (see also \S\ref{s:HDsims}).


In the last decade, more and more cosmological simulation codes have combined chemical enrichment models with many other important physical processes describing the evolution of gas and stars, with the principal aim of reaching an increasingly detailed and realistic picture of cosmic structures, from galaxies to galaxy clusters and cosmic filaments, to be compared against observational findings~\cite[][]{Fabjan:2010,Biffi:2017,Truong:2019,Planelles:2014,Martizzi:2016,Dolag:2017,Barnes:2017,Vogelsberger:2018}.

Nonetheless, most of the numerical studies based on cosmological simulations so far
have concentrated on the case of (massive) galaxy clusters, for which the impact of resolution, feedback processes and interplay with the member galaxy population can be better constrained. 
Given their special position at the crossroad between smaller-scale physics and cosmic evolution, galaxy groups represent in fact a rather challenging, albeit crucial, target: capturing correctly the effects of feedback from central galaxies and BHs, given the shallower potential wells of groups compared to clusters, is of great importance.
This has been in fact the main source of discrepancy in the comparison with observational findings, and consequently one of the crucial testbeds for cosmological simulations and the physical models therein included 
\citep[for a thorough discussion, see the companion review by][]{Oppenheimer:2021}.

\subsubsection{Results from cosmological simulations}\label{s:cosmo-res}

Cosmological simulations can be extremely powerful resources to study and predict the detailed enrichment history of the gas in cosmic structures, as well as to investigate the expected spatial distribution of metals.
It is therefore crucial to assess their reliability by comparing simulated results to observational findings.

As a consequence of the interplay of different physical and dynamical processes, especially energetic feedback, simulations allow us to explore the expected observable signatures on the resulting distribution of metals in the IGrM gas.
Numerical investigations showed, for instance, that feedback from AGNs is crucial to reproduce the large-scale homogeneous enrichment observed in the outer periphery of galaxy systems, such as groups and clusters~\cite[][]{Biffi:2018}.
In such simulations, metallicity profiles show a relatively flat trend out to large distances from the center in clusters, and a very similar enrichment level in smaller structures as well. 
Consistent findings emerge from recent observational studies, as discussed in detail in \S\ref{s:radial}~\cite[see also the review by][]{Mernier:2018Review}.

This effect of early AGN feedback promoting a more homogeneous enrichment and shallower radial profiles, was already observed in the simulations by~\cite{Fabjan:2010}, for both
massive clusters and lower temperature systems. 
At the group regime ($T\lesssim 3$\,keV), they found that the flat profile of iron abundance in the outskirts of the simulated groups was in contrast with the observational results by~\cite{Rasmussen:2007}, employed for comparison~\cite[see also][]{Planelles:2014}. 
The simulation results are instead more in line with recent observational data, e.g.\ by~\cite{Mernier:2017} and~\cite{Lovisari:2019}.
Already in \cite{Fabjan:2010}, it was shown that also the 
the silicon-to-iron abundance ratio in simulations was found to be flat out to large distances from the center in groups, as well as in clusters, indicating also a similar contribution of SNIa and SNcc to the gas metal enrichment. 
A relatively flat silicon-to-iron ratio was also confirmed by independent results obtained by~\cite{McCarthy:2010} on a set of simulated galaxy groups extracted from the OverWhelmingly Large Simulations (OWLS) project~\cite[][]{Schaye:2010}, despite finding typically more pronounced radial gradients of the IGrM metallicity.
Several independent simulation studies converge on the idea that energy feedback solely due to supernova winds typically produces clumpier distributions of $\alpha$-elements, like oxygen or silicon, and steeper, decreasing radial metallicity profiles. 
The reason for the different metal distribution can be related to the origin of the $\alpha$-elements, mostly produced by SNcc and therefore confined in the vicinity of star-formation sites, where they can be efficiently locked back into newly formed stars unless an efficient mechanism intervenes to fastly distribute them far enough.
The steeper profiles in absence of AGN feedback was also mildly noted in~\cite{Dave:2008}, although the galactic outflow model used in their cosmological simulations was able to reproduce the global iron content in group-size haloes, together with various observations of cosmic chemical enrichment. Those authors concluded as well that an efficient outflow mechanism, able to displace pre-enriched gas out of galaxies, must be in place already at early times, in order to explain the observed chemical enrichment of the inter-galactic medium at $z\sim6$~\cite[see also][]{Oppenheimer:2006}.

These general trends have been also confirmed by following simulation campaigns~\cite[][]{Rasia:2015,Biffi:2017,Vogelsberger:2018}. 
In addition to consistent results on the relation between AGN feedback effects and the enrichment level at large distances from the center (i.e.\ beyond $\sim 0.3\,R_{500}$), these recent simulations also finally reproduced the diversity of thermal and chemical properties found in the core ($\lesssim 0.1\times R_{500}$) of observed systems~\cite[][]{Rasia:2015,Barnes:2017}.
Some differences, e.g.\ on the metallicity profile normalization and thus on the global enrichment level, is nonetheless still present depending on the specific set of simulations analysed~\cite[][]{Martizzi:2016}, and consequently on the set of stellar yields and supernova rates adopted.
In addition, the modelling of dust and metal spreading within cosmological simulations can further impact the details of the spatial distribution of metals that remain in the gaseous phase, for which further dedicated studies are needed.

When comparing simulation results with observations, it is important to pay attention to the way quantities are evaluated in simulations. In particular, 
estimates of metallicity (as well as other thermal properties) can be derived in different ways depending on the weight $w$ used to compute the average $Z$ value, that is $Z_w = \int w Z {\rm d}V/\int w {\rm d}V$. Typical weights are the mass of the gas or its emissivity in the X-ray band.
In general, flatter metallicity profiles in simulations are better reproduced when a projected emission-weighted estimate is pursued, whereas the mass-weighted three-dimensional metallicity typically shows a somewhat steeper decrease with radius~\cite[][]{Biffi:2018}.
In this perspective, the issue of a fair comparison between gas properties in simulations and observations has also been tackled via the generation of detailed X-ray synthetic observations out of numerical simulations, properly taking into account the specific characteristics of X-ray telescopes.
With such techniques, \cite{Rasia:2008} showed that an observational-like reconstruction of the iron and oxygen abundances with mock \XMM\ observations of simulated clusters and groups is in good agreement with the intrinsic simulation value.
More recently, \cite{Cucchetti:2018} 
employed synthetic observations of the X-ray Integral Field Unit (X-IFU) on board the next-generation European X-ray observatory {\rm Athena} to reconstruct chemical properties of the ICM from simulated galaxy clusters. The authors showed that the metallicity values obtained from X-IFU spectra match well the emission-measure-estimate computed directly from the simulations.

The homogeneous enrichment of the intra-cluster and intra-group gas on large scales, as indicated by the little scatter around very shallow radial profiles in the outskirts, is also strongly connected to the history of the chemical enrichment. 
The so-called pre-enrichment scenario implies that the gas enriched within proto-groups and proto-cluster galaxies has been displaced beyond their shallower potential wells by some efficient mechanism at early times ($z\gtrsim 3$) --- while stellar feedback alone is not sufficient, many recent state-of-the-art cosmological simulations identify the responsible mechanism with early AGN feedback.
This allows the bulk of the diffuse inter-galactic gas to be pre-enriched and then re-accreted into the assembling galaxy group or cluster. This further supports the idea that a significant fraction of the gas chemical enrichment happens at those early times, as confirmed by the little evolution of the gas metallicity below $z\sim2$, especially on the large scales, found 
in observations~\cite[e.g.][]{Anderson:2009,McDonald:2016,Mantz:2017} and also in simulation studies~\cite[][]{Biffi:2017,Biffi:2018,Vogelsberger:2018}.
At the group scale, cosmological simulations predict a gas metallicity evolution below $z\lesssim1$--$2$ that is very similar to the one found, and observed, in more massive clusters.
Consistently with the results discussed above, simulations including AGN feedback find a shallow dependence on redshift, especially when the global metallicity within $R_{500}$ or the enrichment level in the outer regions {\bf ($\gtrsim 0.3\times R_{500}$)} is concerned. \cite{Truong:2019} show that the evolution of the metallicity in different radial range is similarly mild at groups scales as well, unless only stellar feedback is included in the simulations. In that case, they note again an effective reduction of the gas iron and oxygen content which is more severe particularly in the group regime. This is interpreted as a more substantial, un-suppressed, star formation activity which efficiently, and preferentially, consumes metal-rich gas. In the galactic-outflow model by~\cite{Dave:2008} a higher growth rate, with respect to observations, was in fact observed in the simulated groups.


\begin{figure}
    \centering
    \includegraphics[width=0.7\textwidth]{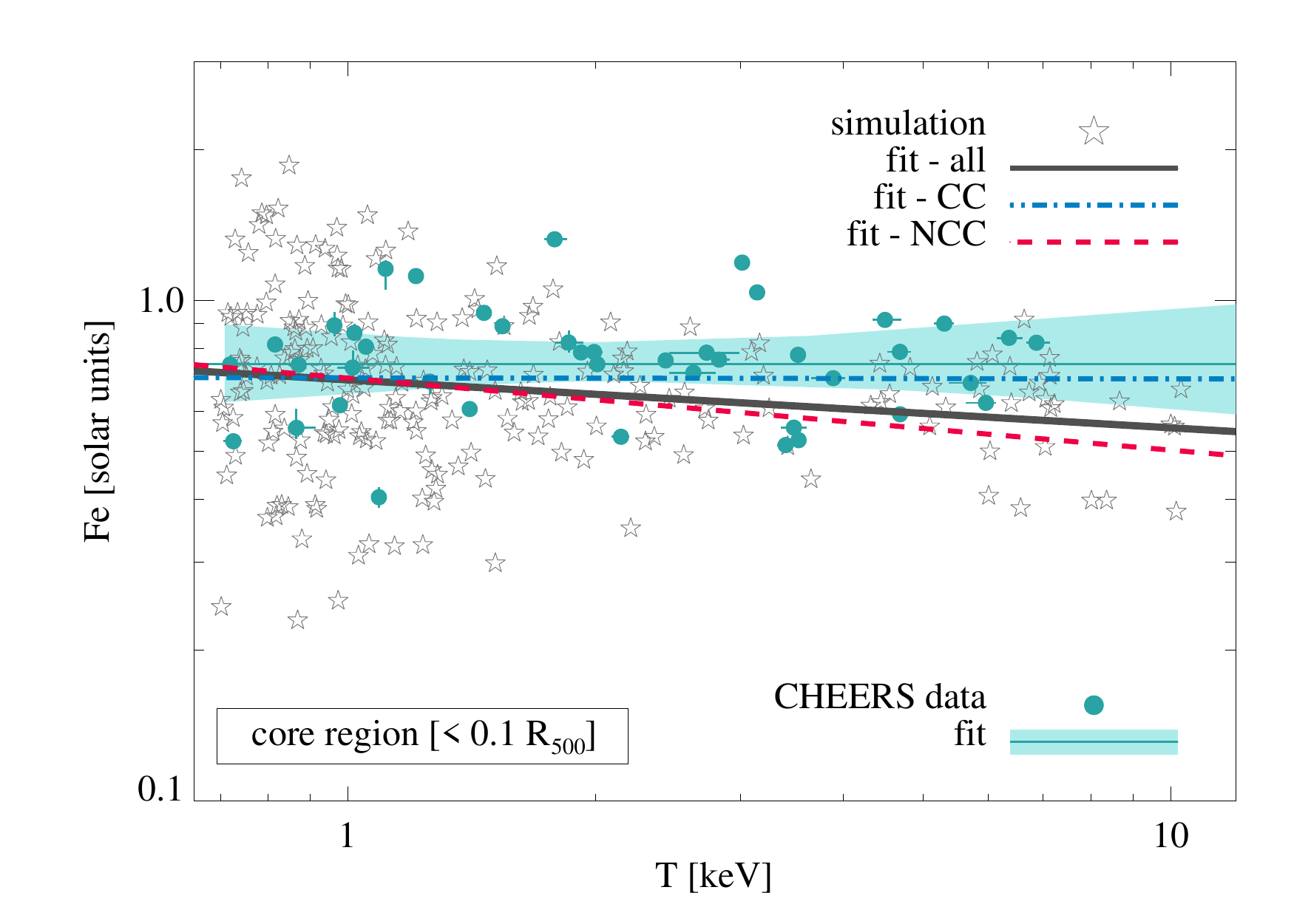}
    \caption{Relation between gas iron abundance and temperature in the core ($r<0.1R_{500}$) of groups and clusters. Comparison between cosmological simulations (empty stars) and X-ray observational results from the CHEERS sample (filled circles). We also report best-fit relations for the whole simulated sample (solid grey line), and for the CC and NCC subsamples (blue dot-dashed and red dashed lines, respectively), as well as the relation determined from the CHEERS data (turquoise line and shaded area, for the associated $68.3\%$ confidence region).
    Consistently with previous Sections, iron abundances, relative to hydrogen, are reported with respect to the Solar reference value by~\cite{Asplund:2009}.
    Adapted from~\protect\cite{Truong:2019}.
    \label{fig:simsT19}}
\end{figure}

The similarities between chemical enrichment of the IGrM in lower-temperature groups and the ICM in massive objects is further supported by the shallow dependence of gas global metallicity on the system mass (or temperature).
In contrast to observational findings, where lower iron abundances were typically observed in group-size systems compared to clusters, \cite{Yates:2017} report
a shallow, mildly anti-correlating, metallicity-temperature relation, employing semi-analytic models of galaxy evolution.
Simulation results, like those presented in~\cite{Dolag:2017} and~\cite{Vogelsberger:2018}, also predict a shallow anti-relation between metallicity and temperature, that extends without breaks from clusters down to groups.
More recently, \cite{Truong:2019}
compare the relation between temperature and iron abundance in the core (i.e.\ $<0.1\,R_{500}$) of simulated groups and clusters with recent results from the CHEERS sample~\cite{Mernier:2018a}, 
also finding a shallow anti-correlation overall, with a mass-invariance of the IGrM and ICM iron abundance in cool-core clusters, as shown in Fig.~\ref{fig:simsT19}.
In the Figure, in particular, the simulated data (star symbols) are reported together with the best-fit relation for the whole sample, and for the CC and NCC subsamples, with the former in better agreement with the CHEERS results~\cite[see also][]{Mernier:2018Review}.
\cite{Dolag:2017} show that a flat relation with temperature applies as well when abundance ratios relative to iron are investigated (e.g. O/Fe, Si/Fe, S/Fe etc.).
Interestingly, this is also in line with recent observational findings~\cite[e.g.][see also discussion in \S\ref{s:composition}]{Mernier:2018b}.

So far, the main limitation to study simulated groups of galaxies in more detail has been the lack of simulations within cosmological context able to consistently resolve and match the stellar and gas properties of massive galaxy clusters and galaxies,
simultaneously. Groups, in this respect, have often rather served as crucial testbed for assessing the prediction power of physical models embedded into cosmological simulations, especially related to feedback from central galaxies (see the companion reviews by \citet{Oppenheimer:2021} and \citet{Eckert:2021}).
Nonetheless, given the recent encouraging results obtained on the chemical and thermal properties of the diffuse gas in various cosmic structures, especially with improvements in the description of chemo-energetic feedback from galaxies in group and cluster cores,
more dedicated studies specifically focusing on the group regime in cosmological simulations are definitely needed to explore the details of their formation and evolution.

\section{Future X-ray missions} \label{s:tele}

Beyond the impressive observational efforts provided by the community to best characterise metals in the IGrM using the past and current fleet of X-ray observatories (Sect. \ref{s:obs_hist} and \ref{s:spatial_distr}), 
it is clear that the next generation of X-ray missions is essential to overcome the current limitations and to advance our knowledge. As we further discuss in this section, higher spectral resolution, higher throughput and larger sky coverage will be key factors.

\subsection{eROSITA}

The present knowledge of the physical properties of the IGrM, both thermodynamical and chemical, is limited to
archival studies with known systems mainly at the high end of the mass and luminosity ranges. X-ray selection provides a more reliable way than optical selection of identifying virialized groups with a bona fide IGrM, but groups are typically at the limit sensitivity of the
\ROSAT\ All Sky Survey (RASS) and their detection is biased towards peaked surface brightness objects \citep[the X-ray cool-core bias,][]{Eckert:2011,Rossetti:2017}. Even though the advent of \XMM\ and \Chandra\ made available 
deep surveys of limited areas providing less biased samples of groups, these systems are typically at moderate redshifts and they lack even global abundance measurements \citep[e.g.][]{Adami:2018,Gozaliasl:2019}. Optically
selected systems with a dedicated X-ray follow-up can circumvent some of the biases of the RASS X-ray selection
and provide some partial answer to the characteristic of the general population of groups \citep[see for example][and the interesting discussion therein]{O'Sullivan:2017}.

\textit{Spectrum Roentgen Gamma} (SRG, launched in 2019) hosts the eROSITA instrument, a set of seven co-aligned soft X-ray telescopes covering the 0.2-10~keV band, with a field of view of 1$^\circ$ and $\sim$15$^{\prime\prime}$ spatial resolution equipped with CCDs with spectral resolution of 60-80 eV in the 0.5-2 keV band \citep{Predehl:2021}. At the end of 2023, after four years surveying the whole sky once every 6 months, eROSITA will build up an all-sky survey 25 $\times$ deeper than RASS in the 0.5-2~keV band \citep{Merloni:2012,Predehl:2021}. 

eROSITA will therefore provide for the first time a large, homogeneously selected X-ray sample of groups for detailed studies with future generations of X-ray instruments \citep[][and see also the companion reviews in particular \citet{Eckert:2021}]{Merloni:2012,O'Sullivan:2017,Kafer:2020} detailed in the next sections and for the pointed phase of eROSITA itself, in a similar fashion as \ROSAT. In particular, given the large field of view, pointed observations of eROSITA will provide valuable information for the outer regions of groups and their metallicity.
These upcoming data sets are all the more interesting in the context of the expected spectral model improvements driven by microcalorimeter data, which will further ensure that the results derived from lower-spectral resolution CCD observations are robust.


\subsection{XRISM} \label{s:XRISM}

\begin{figure}
    \centering
    \includegraphics[width=\textwidth]{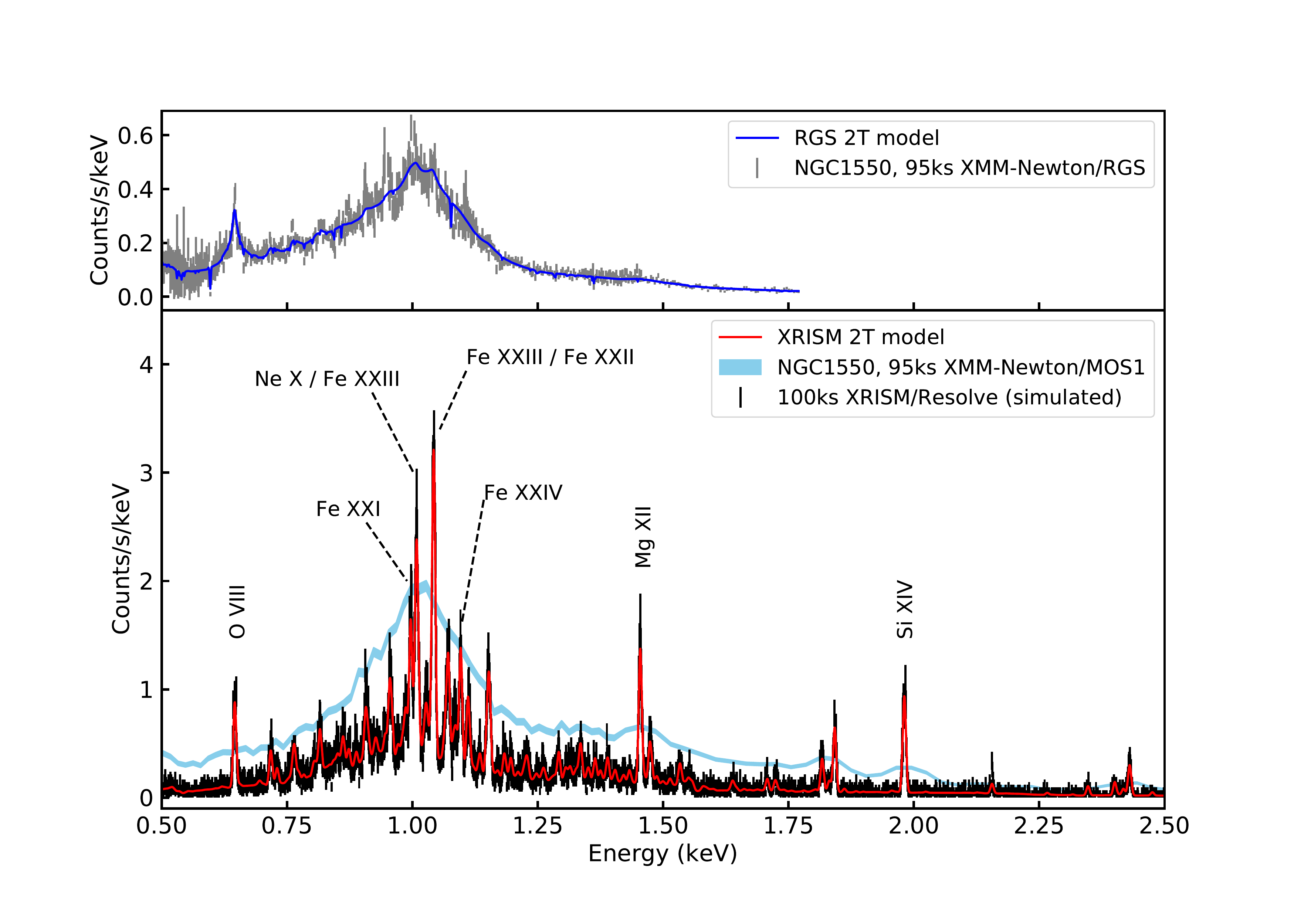}
    \caption{A comparison between existing archival \XMM\ data of the central regions of the galaxy group NGC1550, and predictions for $XRISM/Resolve$ using a similar exposure time and spectral model. \textit{Top panel:} the RGS spectrum extracted from a 3.4 arcmin-wide stripe in the cross-dispersion direction; spectra from RGS1 and RGS2 and from the three different existing observations have been stacked for display purposes \citep[for details of the data reduction, see][]{Pinto:2015}. The blue curve shows the best-fit 2T model using SPEXACT v3.0.6., including the effect of line broadening due to the extent of the source. \textit{Bottom panel:} The \XMM-MOS1 spectrum from the central $0.05\:r_{500}$ region of NGC1550 \citep[adapted from][]{Mernier:2018b} is shown in blue. This region roughly corresponds to the FoV of $XRISM/Resolve$. In red, we show the predicted model obtained by rescaling the 2T RGS model from the top panel to match the \XMM-MOS1 flux, and folding this through the $XRISM/Resolve$ response. This rescaling is necessary because the absolute flux determined by \XMM-RGS for an extended source is uncertain, since technically a region up to 10 arcmin along the dispersion direction can contribute to the observed count rate. A simulated 100~ks $XRISM/Resolve$ spectrum using this model is shown in black. The 5 brightest Fe lines, and lines from all elements other than Fe with a line flux exceeding $5\times10^{-17}$ photons/s/cm$^3$, are labeled.}
    \label{fig:xrism_ngc1550}
\end{figure}

Non-dispersive, high-spectral resolution X-ray micro-calorimeters will enable a giant leap in our understanding of the metal content of the diffuse IGrM. 
The next such detector slated for launch is the $Resolve$ instrument onboard the X-ray Imaging and Spectroscopy Mission ($XRISM$), a JAXA-led satellite with contributions from NASA and ESA \citep{Tashiro:2020} with a launch expected around 2023. The $XRISM/Resolve$ instrument, covering a $3\times3$ arcmin field of view with 35 micro-calorimeter pixels, will carry forward the seminal observations begun by the $Hitomi/SXS$ on the ICM of the Perseus Cluster \citep{Hitomi:2017}. With each pixel delivering a spectral resolution of $<$7~eV, more than 10 times better than conventional CCDs, and given its non-dispersive nature meaning that the spectra of extended sources are not blurred by the size of the target, $XRISM/Resolve$ will reveal emission lines from various chemical elements in the IGrM with unprecedented sharpness (see for example Figure \ref{fig:xrism_ngc1550}). Due to the relatively modest spatial resolution (with a HPD of 1.7 arcmin) and effective area, $XRISM$ observations are ideally suited for studying the centers of nearby clusters and groups. Using these targets, together with dedicated laboratory measurements driven by these new observations \citep[e.g.][]{Betancourt2019,Gu:2020}, it is expected that remaining uncertainties and differences between AtomDB and SPEXACT in modeling the Fe-L line emission will be ironed out early during the lifetime of the mission, whose expected launch is currently set for Japanese fiscal year 2022.

For the bright, line-rich cores of galaxy groups, 100~ks observations with $XRISM/Resolve$ can determine the abundances of Fe, O, Ne, Mg, Si, and S with statistical precisions of 5\% and systematic uncertainties that are far reduced compared to those from fitting CCD spectra. Weaker lines from other elements like N, Al, Ar, Ca, and Ni (based on the Ni L-shell emission) may also be detected in the IGrM. This will provide a fantastic and stringent test of the current picture that the chemical composition of the ICM/IGrM is consistent with the Solar values (see Section \ref{s:composition}). 

Furthermore, precise measurements of the metal abundance patterns of the ICM and IGrM are expected to offer new tests of stellar astrophysics. Elemental abundances in the Sun and in the stars in the Local Group have so far been the most commonly used points of comparison in order to check whether or not current theoretical nucleosynthesis yields provide a self-consistent picture that can appropriately describe Galactic chemical evolution \citep[see e.g.][]{Griffith:2019,Kobayashi:2020}. High-resolution X-ray spectroscopy with $XRISM$, followed by the \Athena\ X-IFU (see Section \ref{s:Athena}) will provide complementary measurements of the enrichment history of the hot-diffuse gas, with a precision rivalling optical and infrared stellar spectroscopy, ushering in a new era of extra-galactic archeology.


 

\subsection{Athena} \label{s:Athena}

In the continuation of the high resolution spectroscopy era to be settled by \XRISM, the future European mission \Athena\ will be a game changer for our understanding of the chemical enrichment of hot haloes pervading large-scale structures, from individual galaxies to rich clusters. Currently planned to be launched around 2033, \Athena\ will embark two revolutionary instruments: the Wide Field Imager \citep[WFI;][]{Rau:2013,Rau:2016} and the X-ray Integral Field Unit \citep[X-IFU;][]{Barret:2013,Barret:2018}. The former consists of a DEPFET (depleted p-channel field-effect transistor) camera able to perform imaging and moderate-resolution spectroscopy over an impressive field of view of $40' \times 40'$, whereas the latter is a cryogenic spectrometer made of a $\sim$5$'$ diameter array of more than 3000 TES (transition edge sensors) -- each of which offering an exquisite spectral resolution of 2.5 eV over a required spatial resolution of 5 arcsec half energy width (thus allowing to probe the spatial substructure if the IGrM at levels comparable to \XMM). Concretely, both instruments will be nicely complementary. While the WFI is expected to discover a large number of high-redshift clusters and groups, the X-IFU will be able to investigate them spectroscopically with unprecedented resolving power. The promises of the latter have been thoroughly demonstrated in the case of clusters, in terms of spatial distribution of metals \citep[][see also Sect. \ref{s:cosmo-res}]{Cucchetti:2018} but also of chemical composition and underlying stellar sources \citep{Mernier:2020b}.

 \begin{figure}[h]
 \centering
     \includegraphics[width=\textwidth, trim={1.2cm 0cm 1.6cm 1cm},clip]{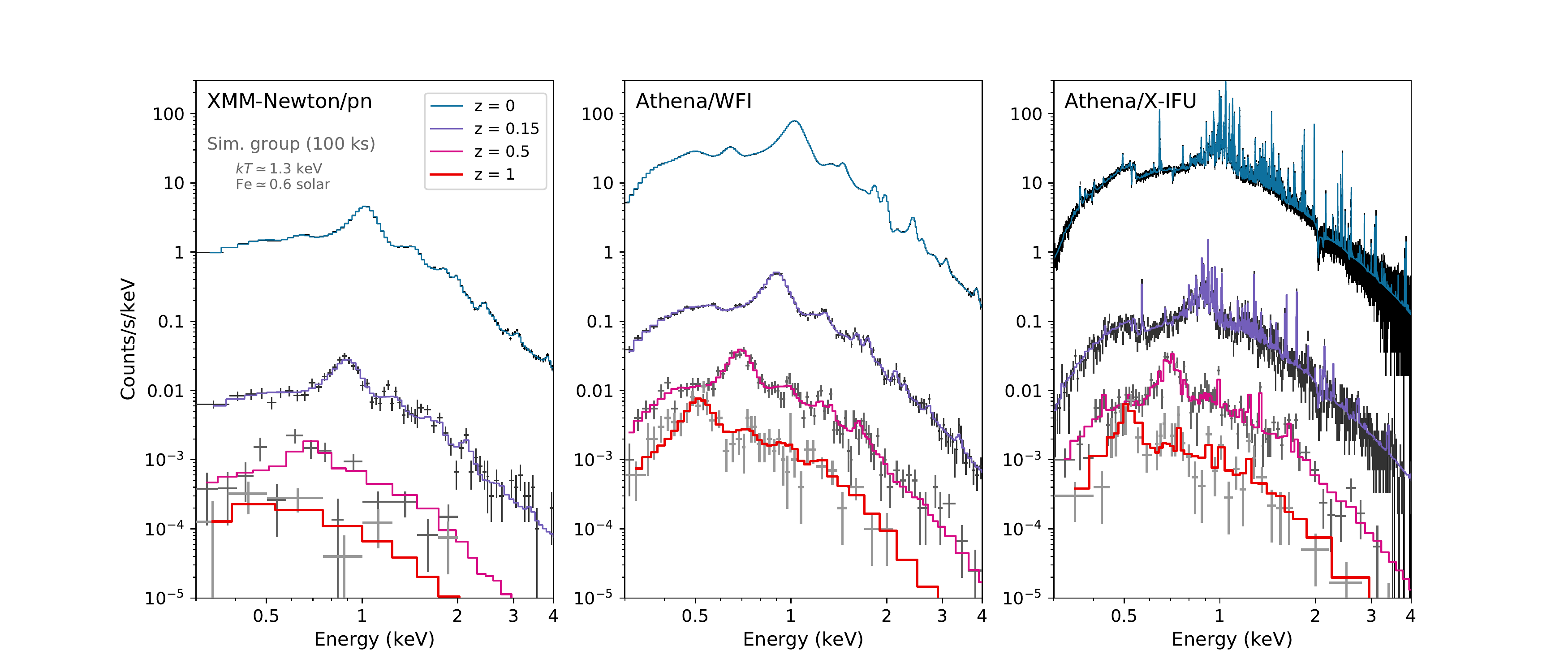}
     \caption{Simulated (100~ks) spectra of the core of a typical NGC\,1550-like group ($L_{0.3-2 \mathrm{keV}} \simeq 2.1 \times 10^{42}$ erg/s, $kT \simeq 1.3 keV$, Fe $\simeq$ 0.6 Solar) set at various redshifts, as seen by the \XMM/pn, \Athena/WFI, and \Athena/X-IFU instruments. Each spectrum has been appropriately re-binned for clarity.}
     \label{fig_Athena}
 \end{figure}

The outstanding scientific potential offered by \Athena\ for exploring the metal content of the IGrM is illustrated in Fig.~\ref{fig_Athena}, where we simulated 100~ks of WFI and X-IFU exposures (in comparison to that of \XMM/pn) for the core ($<$0.05 $r_{500}$) of a typical NGC\,1550-like group assumed at various redshifts. Whereas for this specific case pn cannot provide significant metal constraints at $z = 0.5$ and beyond, we calculate that the WFI and the X-IFU should be able to track the overall metallicity up to $z = 1$ within $\sim$40\% or less. An interesting feature immediately visible from this figure is the possibility of performing spectroscopy on high-redshift groups with the WFI. As the effective spectral resolution of any instrument naturally tends to deteriorate with decreasing flux (in order to keep enough meaningful statistics per spectral bin), at $z = 1$ typical WFI and X-IFU spectra of groups are expected to deliver very similar spectral information. Combined to its ability to detect and image several groups simultaneously over a large region of the sky \citep[more than 10,000 systems at $z > 0.5$ with $M \ge 10^{13} M_\odot$ over the nominal four years of the mission;][]{Zhang:2020}, this will make the WFI a highly valuable instrument to trace the metal content of \textit{many} (local and distant) groups at once. This actually provides a unique synergy with the X-IFU, as the latter will be invaluable to resolve a plethora of metal lines at low and moderate redshifts (in order to derive absolute and relative abundances with exquisite accuracy, but also to further refine atomic calculations and make the available spectral codes further converge).

Another particularly interesting possibility offered by the X-IFU instrument resides in the determination of the abundances ratios. 
By giving us access to even more metals with even fainter lines than $XRISM$, the X-IFU will offer the best diagnostics of the SN explosion mechanisms, initial metallicity of the progenitor stars contributing to the enrichment of the cosmos, and especially the slope of the IMF, with the aim to contribute in a significant way to the debate about its universality. 
Higher line emissivities of a few key elements (e.g. O, Mg) at the groups regime are critical for constraining the IMF and will nicely complement the case of clusters, which will be thoroughly studied as well \citep[see e.g.][]{Mernier:2020b}.

Although more tailored WFI and X-IFU predictions (including e.g. cosmological evolution of groups, the effects of the background and its reproducibility and/or other instrumental effects) are left for future dedicated work, it is clear that \Athena\  will push our understanding of the chemistry of large-scale structures to the next level, even at and below groups scales. 

\subsection{HUBS and Super DIOS}

\begin{figure}[t]
    \centering
    \includegraphics[width=0.65\textwidth]{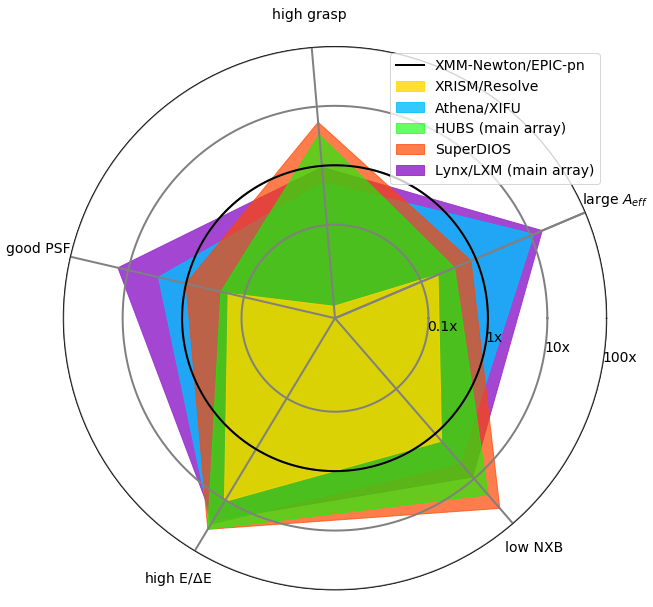}
    \caption{Summary of the capabilities of future missions and mission concepts carrying high-spectral resolution X-ray integral field unit detectors, using the \textit{XMM-Newton} EPIC/pn as a reference value of 1 along each axis.
   The grasp is defined as the effective area $A_{eff}$ integrated over the field of view; plotted along the `low NXB' axis are the non-dimensional figures of merit $F_{NXB}$ for detector-background-limited observations (also known as non-X-ray background), expressed as $F_{NXB}=A_{eff}/(F^2B)$ with F the mirror focal length and B=1 in low-Earth orbit and 4 for high orbits, as defined by e.g. \cite{Mushotzky:2019}.}
    \label{fig:future}
\end{figure}

While $XRISM$ and $Athena$ will lead to significant advancements in our understanding of the precise chemical make-up of the centers of galaxy groups and its redshift evolution, the high spectral resolution IFUs onboard both of these future missions have a limited field of view. This means that studies of the metal abundance ratios in the outskirts of groups and clusters would be very expensive in terms of observing time: nearby objects would be too large on the sky, and require mosaics composed of an unwieldy number of pointings, while high-redshift objects whose outskirts do fit within the FoV would be significantly dimmer. To give a concrete example, covering the entire area between 1--1.5 $R_{500}$ of NGC~5846\footnote{for the $R_{500}$ value in  \cite{Panagoulia:2014_prof}} would require no less than 423 observations with the $Athena$/X-IFU. 

Two future missions currently under study promise to address this issue by offering capabilities that are complementary to those of $Athena$: the Hot Universe Baryon Surveyor ({\it HUBS}) \citep{Cui:2020}, which is a project of the Chinese National Space Administration (CNSA), and the JAXA-led {\it Super DIOS} ("Diffuse Intergalactic Oxygen Surveyor", \citealt{Sato:2020}) mission. Both have expected launch dates in the 2030s. 
By prioritising a shorter mirror focal length (which implies a smaller on-axis effective area) but using larger pixels and covering a larger field of view (of order $\sim$1 deg$^2$), these future satellites will be able to survey a wider area of the sky more efficiently while maintaining a high spectral resolution ($\Delta E \lesssim2$~eV). Due to the shorter focal length and their planned deployment in low-Earth orbit which both help to minimise the detector background, these missions are optimised to enable detailed high-resolution spectroscopy of the faint outskirts of nearby groups and clusters of galaxies, among several other science topics. In Fig. \ref{fig:future}, we summarise the capabilities and advantages of planned X-ray IFUs over the next $\sim$decade. Remarkable to note is that the {\it Super DIOS} concept plans to employ nearly an order of magnitude more TES pixels than $Athena$ with developments of a new TES readout system, enabling a 10--15 arcsec resolution over 0.5-1 deg$^2$; while HUBS has a similar number of pixels as $Athena$ (and therefore a poorer spatial resolution of $\sim$1 arcmin over a 1 deg$^2$ FoV), it carries a central 12 by 12 arcmin sub-array with an energy resolution of 0.6~eV, that will be the first detector to reach a resolving power of $E/\Delta E>1000$ around the Fe-L complex.


\subsection{Arcus}

Arcus is a proposed NASA Medium Explorer mission that aims
to significantly improve our spectroscopic capabilities using soft X-ray gratings. 
With spectral resolution $R>$2500 between 12-50 $\mathring{A}$, the
Arcus proposal possesses the ability to resolve the absorption of a
diverse range of metal species in the extended halos of galaxies,
groups, and clusters along sight lines toward distant quasars
\citep{smith16}.  X-ray grating spectroscopy provides a critical
complementary probe to planned micro-calorimeters by having $\sim
10\times$ greater resolution and probing the diffuse hot gas
comprising the majority of missing metals and baryons
\citep{cen06,wijers19}.  Arcus is designed to achieve a 10-fold
increase in sensitivity (i.e. square root of effective area times
resolution) over existing grating spectrometers on \Chandra\ and \XMM.

According to cosmological simulations \citep{wijers20}, the IGrM has
the greatest potential to show a rich variety of ion species.  O VIII
should be the most detected species within group virial radii, while O
VII should be detected in lower mass groups at $M_{500}\la 10^{13.5}
M_{\odot}$.  In its main mission, Arcus should detect the
$T=10^{5.4-6.8}$ K IGrM in at least 10 group haloes at a 3 m$\mathring{A}$
equivalent width threshold (at $5\sigma$) for O VII and O VIII.
Interestingly, the Fe XVII should be detectable within $R_{500}$ for
groups, probing gas up to $\approx 10^7$ K for at least 5 group
haloes.  Integrating longer on the brightest X-ray quasars opens up
the possibility to detect more species using a 1 or 2 m$\mathring{A}$ detection
threshold, including Ne IX and Ne X.  C V and C VI (and maybe N VII)
may also be detectable in the group outskirts for deeper observations
probing gas down to $10^5$ K.  Sensitive UV absorption line
spectroscopy brought about by the Cosmic Origins Spectrograph on {\it
Hubble} discovered that the objects with the richest and most diverse
set of species probing gaseous halos are star-forming, $L^*$ galaxies
\citep{werk14}.  It is expected that groups will be the analogous
richly detected, multi-species objects to be probed via X-ray
absorption line spectroscopy.

\subsection{Lynx}

The $Lynx$ mission concept \citep{Ozel:2018} proposed to the US 2020 Decadal Survey is designed to carry a new generation of X-ray telescopes enabling a sub-arcsecond resolution over a 22$^\prime\times$22$^\prime$ field of view and an effective area of 2~m$^2$ at 1~keV. It is designed to be equipped with three complementary instruments. An active pixel array (the high-definition X-ray imager, HDXI) that would provide wide-field CCD-like spectral imaging, but with 0.3$^{\prime\prime}$ pixels to take advantage of the exquisite spatial resolution. The $Lynx$ X-ray Microcalorimeter (LXM) which would bring 3~eV spectral resolution on 1$^{\prime\prime}$ spatial scales over a 5$^\prime$ field of view (0.3 eV in a ultra-high resolution sub-array). Finally an X-ray grating spectrometer (XRG) with an effective area of 4000 cm$^2$ and resolving power greater than 5000 to exploit the better spectral resolution of gratings for point sources in the soft energy band.
The key improvements $Lynx$ will bring if approved concern the ability to resolve the metal abundance structure of the gas near and outside the virial radius of groups at low redshift by combining a study of emission and absorption against background AGNs. $Lynx$ will also push the realm of enrichment studies in the IGrM to the z=2-3 range where strong trends are expected. $Lynx$ is specifically targeting observations of high-redshift groups down to a mass scale of $M_{500} = 2 \times 10^{13}$ $\msun$ at z $>3$ \citep[\url{https://www.lynxobservatory.com/report}, ][]{Gaskin:2019}.

\section{Concluding Remarks}\label{s:conclusions}

Throughout this review, we have seen how crucial the case of galaxy groups is in order to complete our understanding of the journey of metals -- from stars and supernovae to the largest scales of our Universe. Groups are in fact a unique piece of the puzzle to relate chemical enrichment at (sub-) galactic scales and at cluster scales (Sect. \ref{s:mlr}), making IGrM abundance studies particularly valuable in this respect. As we have also discussed however, such studies are still in their pathfinder steps, and enrichment in groups remains much less explored (hence, understood) than in clusters. The reasons are diverse, and include notably on the observational side: (i) the intrinsic faintness of the IGrM with respect to the ICM, making observational studies more demanding in terms of exposure time; (ii) the lack of well defined samples of groups in the X-ray band, marking again a stark contrast with respect to clusters; (iii) our lack of spectral knowledge of the Fe-L complex (ruling almost all the X-ray emission of the IGrM) and of the likely multi-temperature structure of the gas (Sect. \ref{s:Fe-L}). 
As we have seen in Sect.~\ref{s:tele}, in the next years each of the above limitations will be tackled by higher quality data in terms of sizes of the sample of groups to be targeted and available high-throughput and high-resolution spectra. Those data will be coupled with an improved theoretical understanding of the spectral modeling. It is also essential to continue the efforts to improve the accuracy of the theoretical SN yields and the measurements of SN rates as a function of the cosmic time.
On the simulation side, the main challenge will be to reproduce key stellar and gaseous properties on both galactic and Mpc scales, simultaneously including micro-scale physics and the large-scale cosmological context (Sect.~\ref{s:HDsims} and \ref{s:cosmosims}). With the ongoing exponential high-performance computing advancements, we are getting closer to a quantum leap in terms of a single high-resolution hydrodynamical cosmological simulation reaching this goal, thus providing more detailed predictions for the regime of galaxy groups.
It is a reasonable bet to forecast the coming era as a golden age of maturity for 
the study of metal abundances in galaxy groups.

\vspace{6pt} 

\supplementary{The review is based on public data and/or published papers. The results collected from literature and used to generate some of the figures in this review can be found at this link \url{https://doi.org/10.5281/zenodo.5011831}}


\funding{``This research received no external funding''}
\authorcontributions{F.G.: lead author, Sect. 1, 2.2; A.S: Sect. 2.1, 2.3.2, 2.3.3; F.M.: Sect. 2.3.1, 2.5; V.B.: Sect. 3.2; M.G.: Sect. 3.1; K.S and K.M: Sect. 2.4 with contribution from F.G.
F.G., A.S., F.M. and K.S contributed to Sect. 4. All authors contributed to Sect.5. All authors have read and agreed to the published version of the manuscript. }

\acknowledgments{We thank the two anonymous referees for helpful reports which improved the quality of this review. We would like to thank Ben Oppenheimer for providing the section of the future mission Arcus and for useful discussions. We would like to thank David Buote, Silvano Molendi, Simona Ghizzardi, Fabrizio Brighenti, William Mathews for comments and suggestions.
F.G acknowledges financial contribution from INAF "Call per interventi aggiuntivi a sostegno della ricerca di main stream di INAF".
AS is supported by the Women In Science Excel (WISE) programme of the NWO, and acknowledges the World Premier Research Center Initiative (WPI) and the Kavli IPMU for the continued hospitality. SRON Netherlands Institute for Space Research is supported financially by NWO. MG acknowledges partial support by NASA Chandra GO8-19104X/GO9-20114X and HST GO-15890.020-A grants.}

\conflictsofinterest{The authors declare no conflict of interest.} 





\externalbibliography{yes}
\bibliographystyle{Definitions/aa}
\bibliography{metals_groups,biblio2,mlr_ref}

\end{document}